\documentclass[aps,prc,  reprint,secnumarabic, superscriptaddress, nofootinbib]{revtex4-1}
	
\usepackage{graphicx}

\graphicspath{{./Figures/}}  

\begin{document}

 \makeatletter 
 \def\@eqnnum{{\normalsize \normalcolor (\theequation)}} 
  \makeatother


\title{Determination of the Beam-Spin Asymmetry of Deuteron Photodisintegration\\ in the Energy Region $E_\gamma=1.1-2.3$~GeV}



\newcommand*{\ANL}{Argonne National Laboratory, Argonne, Illinois 60439}
\newcommand*{\ANLindex}{1}
\affiliation{\ANL}
\newcommand*{\ASU}{Arizona State University, Tempe, Arizona 85287-1504}
\newcommand*{\ASUindex}{2}
\affiliation{\ASU}
\newcommand*{\CSUDH}{California State University, Dominguez Hills, Carson, CA 90747}
\newcommand*{\CSUDHindex}{3}
\affiliation{\CSUDH}
\newcommand*{\CANISIUS}{Canisius College, Buffalo, NY}
\newcommand*{\CANISIUSindex}{4}
\affiliation{\CANISIUS}
\newcommand*{\CMU}{Carnegie Mellon University, Pittsburgh, Pennsylvania 15213}
\newcommand*{\CMUindex}{5}
\affiliation{\CMU}
\newcommand*{\CUA}{Catholic University of America, Washington, D.C. 20064}
\newcommand*{\CUAindex}{6}
\affiliation{\CUA}
\newcommand*{\SACLAY}{CEA, Centre de Saclay, Irfu/Service de Physique Nucl\'eaire, 91191 Gif-sur-Yvette, France}
\newcommand*{\SACLAYindex}{7}
\affiliation{\SACLAY}
\newcommand*{\UCONN}{University of Connecticut, Storrs, Connecticut 06269}
\newcommand*{\UCONNindex}{8}
\affiliation{\UCONN}
\newcommand*{\FU}{Fairfield University, Fairfield CT 06824}
\newcommand*{\FUindex}{9}
\affiliation{\FU}
\newcommand*{\FIU}{Florida International University, Miami, Florida 33199}
\newcommand*{\FIUindex}{10}
\affiliation{\FIU}
\newcommand*{\FSU}{Florida State University, Tallahassee, Florida 32306}
\newcommand*{\FSUindex}{11}
\affiliation{\FSU}
\newcommand*{\Genova}{Universit$\grave{a}$ di Genova, 16146 Genova, Italy}
\newcommand*{\Genovaindex}{12}
\affiliation{\Genova}
\newcommand*{\GWUI}{The George Washington University, Washington, DC 20052}
\newcommand*{\GWUIindex}{13}
\affiliation{\GWUI}
\newcommand*{\ISU}{Idaho State University, Pocatello, Idaho 83209}
\newcommand*{\ISUindex}{14}
\affiliation{\ISU}
\newcommand*{\INFNFE}{INFN, Sezione di Ferrara, 44100 Ferrara, Italy}
\newcommand*{\INFNFEindex}{15}
\affiliation{\INFNFE}
\newcommand*{\INFNFR}{INFN, Laboratori Nazionali di Frascati, 00044 Frascati, Italy}
\newcommand*{\INFNFRindex}{16}
\affiliation{\INFNFR}
\newcommand*{\INFNGE}{INFN, Sezione di Genova, 16146 Genova, Italy}
\newcommand*{\INFNGEindex}{17}
\affiliation{\INFNGE}
\newcommand*{\INFNRO}{INFN, Sezione di Roma Tor Vergata, 00133 Rome, Italy}
\newcommand*{\INFNROindex}{18}
\affiliation{\INFNRO}
\newcommand*{\INFNTUR}{INFN, Sezione di Torino, 10125 Torino, Italy}
\newcommand*{\INFNTURindex}{19}
\affiliation{\INFNTUR}
\newcommand*{\ORSAY}{Institut de Physique Nucl\'eaire, CNRS/IN2P3 and Universit\'e Paris Sud, Orsay, France}
\newcommand*{\ORSAYindex}{20}
\affiliation{\ORSAY}
\newcommand*{\ITEP}{Institute of Theoretical and Experimental Physics, Moscow, 117259, Russia}
\newcommand*{\ITEPindex}{21}
\affiliation{\ITEP}
\newcommand*{\JMU}{James Madison University, Harrisonburg, Virginia 22807}
\newcommand*{\JMUindex}{22}
\affiliation{\JMU}
\newcommand*{\KNU}{Kyungpook National University, Daegu 702-701, Republic of Korea}
\newcommand*{\KNUindex}{23}
\affiliation{\KNU}
\newcommand*{\UNH}{University of New Hampshire, Durham, New Hampshire 03824-3568}
\newcommand*{\UNHindex}{24}
\affiliation{\UNH}
\newcommand*{\NSU}{Norfolk State University, Norfolk, Virginia 23504}
\newcommand*{\NSUindex}{25}
\affiliation{\NSU}
\newcommand*{\MSUU}{Mississippi State University, MS 39762}
\newcommand*{\MSUUindex}{26}
\affiliation{\MSUU}
\newcommand*{\OHIOU}{Ohio University, Athens, Ohio  45701}
\newcommand*{\OHIOUindex}{27}
\affiliation{\OHIOU}
\newcommand*{\ODU}{Old Dominion University, Norfolk, Virginia 23529}
\newcommand*{\ODUindex}{28}
\affiliation{\ODU}
\newcommand*{\RPI}{Rensselaer Polytechnic Institute, Troy, New York 12180-3590}
\newcommand*{\RPIindex}{29}
\affiliation{\RPI}
\newcommand*{\UOR}{University of Richmond, Richmond, Virginia 23173   }
\newcommand*{\UORindex}{30}
\affiliation{\UOR}
\newcommand*{\ROMAII}{Universita' di Roma Tor Vergata, 00133 Rome Italy}
\newcommand*{\ROMAIIindex}{31}
\affiliation{\ROMAII}
\newcommand*{\MSU}{Skobeltsyn Institute of Nuclear Physics, Lomonosov Moscow State University, 119234 Moscow, Russia}
\newcommand*{\MSUindex}{32}
\affiliation{\MSU}
\newcommand*{\SCAROLINA}{University of South Carolina, Columbia, South Carolina 29208}
\newcommand*{\SCAROLINAindex}{33}
\affiliation{\SCAROLINA}
\newcommand*{\TEMPLE}{Temple University,  Philadelphia, PA 19122 }
\newcommand*{\TEMPLEindex}{34}
\affiliation{\TEMPLE}
\newcommand*{\JLAB}{Thomas Jefferson National Accelerator Facility, Newport News, Virginia 23606}
\newcommand*{\JLABindex}{35}
\affiliation{\JLAB}
\newcommand*{\UTFSM}{Universidad T\'{e}cnica Federico Santa Mar\'{i}a, Casilla 110-V Valpara\'{i}so, Chile}
\newcommand*{\UTFSMindex}{36}
\affiliation{\UTFSM}
\newcommand*{\EDINBURGH}{Edinburgh University, Edinburgh EH9 3JZ, United Kingdom}
\newcommand*{\EDINBURGHindex}{37}
\affiliation{\EDINBURGH}
\newcommand*{\GLASGOW}{University of Glasgow, Glasgow G12 8QQ, United Kingdom}
\newcommand*{\GLASGOWindex}{38}
\affiliation{\GLASGOW}
\newcommand*{\VIRGINIA}{University of Virginia, Charlottesville, Virginia 22901}
\newcommand*{\VIRGINIAindex}{39}
\affiliation{\VIRGINIA}
\newcommand*{\WM}{College of William and Mary, Williamsburg, Virginia 23187-8795}
\newcommand*{\WMindex}{40}
\affiliation{\WM}
\newcommand*{\YEREVAN}{Yerevan Physics Institute, 375036 Yerevan, Armenia}
\newcommand*{\YEREVANindex}{41}
\affiliation{\YEREVAN}

\newcommand*{\NOWODU}{Old Dominion University, Norfolk, Virginia 23529}
\newcommand*{\NOWINFNGE}{INFN, Sezione di Genova, 16146 Genova, Italy}
\author{N. Zachariou}  
\affiliation{\SCAROLINA}
\affiliation{\GWUI}
\author{Y. Ilieva}  
\affiliation{\SCAROLINA}
\author{N. Ya. Ivanov}  
\affiliation{\YEREVAN}
\author{M. M. Sargsian}  
\affiliation{\FIU}
\author{R. Avakian}  
\affiliation{\YEREVAN}
\author{G. Feldman}  
\affiliation{\GWUI}
\author{P. Nadel-Turonski}  
\affiliation{\JLAB}
\affiliation{\GWUI}
\author {K.P. ~Adhikari} 
\affiliation{\ODU}
\author {D.~Adikaram} 
\affiliation{\ODU}
\author {M.D.~Anderson} 
\affiliation{\GLASGOW}
% ...

\collaboration{The CLAS Collaboration}
\noaffiliation


\date{\today}

\begin{abstract}
The beam-spin asymmetry, $\Sigma$, for the reaction $\gamma d\rightarrow pn$ has been measured using the CEBAF Large Acceptance Spectrometer (CLAS) at  the Thomas Jefferson National Accelerator Facility (JLab) for six photon-energy bins between 1.1 and 2.3 GeV, and proton angles in the center-of-mass  frame, $\theta_{c.m.}$, between $25^\circ$ and $160^\circ$. These are the first measurements of beam-spin asymmetries at $\theta_{c.m.}=90^\circ$ for photon-beam energies above 1.6~GeV, and the first measurements for angles other than $\theta_{c.m.}=90^\circ$. The angular and energy dependence of $\Sigma$ is expected to aid in the development of QCD-based models to understand the mechanisms of deuteron photodisintegration in the transition region between hadronic and partonic degrees of freedom, where both effective field theories and perturbative QCD cannot make reliable predictions.
\end{abstract}


\maketitle

\section{Introduction}\label{intro}
\subsection{\emph{Scaling laws in QCD and experimental data}}
The process of  deuteron photodisintegration,
\begin{equation}
\gamma +d\rightarrow n+p,\label{1}
\end{equation}
is especially important for the investigation of the role of quarks and gluons in nuclear
interactions. This photonuclear reaction is: \emph{i)} the simplest ($A=2$) and \emph{ii)} well
studied experimentally. During the last 25 years a number of experiments have measured its differential cross section over a broad range in energy and angle
\cite{Deute01, Deute02, Deute03, Deute04, Deute05, Deute06, Deute06a, Deute09}. There are also some data on the recoil proton polarization~\cite{Deute07, PhysRevLett.98.182302}
and the single beam-spin asymmetry~\cite{Deute11, Deute12}.

The most remarkable property of the available cross-section data is the energy behavior of this
photonuclear process. At photon energies $E_{\gamma}\geq 1$ GeV and large proton scattering angles, it was found that
${\rm d}\sigma/{\rm d}t(s,\theta_{c.m.})\sim s^{-11}$, where $s$ and $t$ (and $u$ referred to later in the paper) are the usual
Mandelstam variables denoting the square of the center-of-mass energy and the square of the four-momentum transfer to the neutron, while $\theta_{c.m.}$ is the proton scattering angle in the center-of-mass frame
(for more details and results on the scaling behavior of the differential cross section, see Ref.~\cite{Deute13}).
Such a behavior is predicted by the constituent counting rules (CCR) based on the scaling
law for  hadron wave functions~\cite{Deute17a,Deute17}. For an arbitrary exclusive two-body reaction at large $s$ and $t$,
CCR predict a power-law falloff of the production cross section at fixed angles:
\begin{equation}
{\rm d}\sigma/{\rm d}t\sim h(\theta_{c.m.})/s^{n-2}, \label{2}
\end{equation}
where $n$ is the total number of elementary fields in the initial and final states,
while $h(\theta_{c.m.})$ depends on details of the dynamics of the process.

The quark counting rule was originally obtained
based on dimensional analysis under the assumptions that the only
scales in the system are momenta and that composite
hadrons can be replaced by point-like constituents with zero angular momentum~\cite{Deute17a,Deute17}.
Later, these counting rules were confirmed within the framework of
perturbative QCD (pQCD) up to logarithmic factors by showing that exclusive two-body reactions at large $s$ and $t$ are dominated by quark and gluon subprocesses at short distances~\cite{Lepage}. 
Within this framework, dimensional scaling can be justified only in the high-energy limit, $t\sim s \gg m^{2}$, where one can neglect the masses, $m$, of the interacting particles. Therefore, one would not expect that the CCR will hold in the few-GeV region.
However, an all-order demonstration of the counting
rules for hard exclusive processes has been shown to arise from the correspondence
between a string theory in anti-de Sitter space and conformal field theories (AdS/CFT)
in physical space-time~\cite{cft1, cft2, cft3}. 
The AdS/CFT correspondence~\cite{maldacena}
leads to an analytical, semi-classical model for strongly-coupled QCD, which has
scale invariance and dimensional counting at short distances, and color confinement
at large distances. In this model, dimensional scaling occurs not only at very large but also at very small 
momentum transfer, $Q$, to the parton. In the latter situation, scaling is due to the constancy of 
the strong coupling with $Q$ when $Q$ is very small.
The AdS/CFT derivation of the scaling laws is particularly interesting since it is a non-perturbative
derivation, which suggests that dimensional scaling is a feature of both perturbative and non-perturbative dynamics.
Experimental studies of nuclear reactions, such as deuteron photodisintegration, 
where the overall momentum transfer is distributed among many constituents, so that
the momentum transfer per parton is small,
are needed to test this model.

An approximate dimensional scaling  has been observed in many exclusive
reactions at sufficiently high energy and large momentum transfer (for reviews see
Refs.~\cite{diff_cross,white, holtgilman}). In addition, the low-energy data on deuteron
photodisintegration~\cite{Deute01, Deute02, Deute03, Deute04, Deute05, Deute06, Deute06a, Deute09} (as well as charged-pion photoproduction
\cite{zhu1, zhu2}) also demonstrate scaling behavior. 
 To understand the observed energy behavior it is useful to look closely  at previous claims of agreement between  data for differential  cross sections and the CCR predictions. 
In fact, the scaled 90$^\circ$
center-of-mass $pp$ elastic scattering data $s^{10}{{{\rm d}\sigma}/{{\rm d}t}}$ show
substantial oscillations about the power-law behavior~\cite{p-p1, p-p1a, p-p1b,p-p2}. Such
oscillations are also seen in $\pi p$ fixed-angle scattering~\cite{pi-p, pi-pa, pi-pb}.
The old data~\cite{diff_cross} as well as the newer data from JLab experiment E94-104
on photoproduction of charged pions at $\theta_{c.m.} = 90^\circ$~\cite{zhu1, zhu2} also show
hints of oscillation about the expected $s^{-7}$ scaling. There are hints of
scaling behavior in
$\gamma d \rightarrow d \pi^0$ as well~\cite{meekins,ilieva} ($s^{-13}$ in this case).
A theoretical interpretation of this oscillatory behavior of the scaled cross section was attempted by many authors, with the more successful interpretations taking  into account the  orbital angular momentum of the partons and hadron helicity flip, relating this oscillatory behavior to spin-dependent effects.

The experimental investigations of  scaling phenomena and related spin-dependent effects resulted in significant theoretical advances in
understanding the role and range of applicability of perturbative QCD at low and
intermediate energies. These studies make it possible to develop a number of nonperturbative
QCD-based approaches to the hadronic dynamics at long distances.
The results achieved to date provide a strong motivation for further investigation
of scaling laws and spin effects in  photonuclear reactions through studies of polarization observables.
Measurements of the beam-spin asymmetry, $\Sigma$,  in deuteron photodisintegration, defined as
{\small
\begin{eqnarray}
\Sigma=&\frac{2Re\left[ \sum_{\pm}\left(F_{1\pm}^*F_{3\mp}^{}-F_{4\pm}^{} F_{6\mp}^*\right)-F_{2+}^*F_{2-}^{}+F_{5+}^*F_{5-}^{}\right]}{f(\theta)},\\
\mbox{where }&f(\theta)=\sum_{i=1}^6\left[|F_{i+}|^2+|F_{i-}|^2\right],\nonumber
\end{eqnarray}} 
and $F_{i\pm}=\langle\lambda_p, \lambda_n|T| \lambda_\gamma, \lambda_d \rangle $ are the helicity amplitudes of the reaction, as defined in Ref.~\cite{HelcAmpl}:
\begin{eqnarray}
&F_{1\pm}=\langle\pm\frac{1}{2}, \pm\frac{1}{2}|T| 1, 1 \rangle,\;\; F_{2\pm}=\langle\pm\frac{1}{2}, \pm\frac{1}{2}|T| 1, 0 \rangle,\nonumber\\
&F_{3\pm}=\langle\pm\frac{1}{2}, \pm\frac{1}{2}|T| 1, -1 \rangle, F_{4\pm}=\langle\pm\frac{1}{2}, \mp\frac{1}{2}|T| 1, 1 \rangle,\nonumber\\
&F_{5\pm}=\langle\pm\frac{1}{2}, \mp\frac{1}{2}|T| 1, 0 \rangle,\;\;\;\;\; F_{6\pm}=\langle\pm\frac{1}{2}, \mp\frac{1}{2}|T| 1, -1 \rangle.\nonumber
\end{eqnarray}
These amplitudes can give access to important aspects of the underlying physics, such as QCD  final-state
interactions and quark orbital angular momentum in the lightest nuclei. The most popular
quark-gluon models for deuteron photodisintegration are the reduced nuclear
amplitudes (RNA) model~\cite{Deute21, Deute21a}, the hard-rescattering  mechanism (HRM)~\cite{Deute24, Deute25, Deute25a, misakgranados},
and the quark-gluon string model (QGSM)~\cite{Deute22, Deute23}.

\subsection{\emph{Theoretical models }}
\subsubsection{Reduced Nuclear Amplitudes Model (RNA)}
The idea of RNA was introduced by Brodsky and
Hiller in order to extend the region of applicability of pQCD down to lower momentum transfers by incorporating some of the soft physics not described by pQCD~\cite{Deute21, Deute21a}. This is done using experimentally determined nucleon form factors to describe the gluon exchanges within the nucleons.
It is hoped that the resulting expressions correctly include
much of the missing soft physics and would therefore be valid for  momentum transfers lower than the  ones in the original pQCD expressions. The RNA calculation is only available at $\theta_{c.m.}=90^\circ$ and makes no predictions for the angular dependence of the cross section. It also does not include spin-dependent effects and thus cannot make predictions for polarization observables.
\subsubsection{Hard-Rescattering Mechanism (HRM)}
In the HRM model it is assumed that large-angle hard breakup of the deuteron is a two-step process~\cite{Deute24, misakgranados}. In the first step, the photon knocks out a quark from one nucleon. Then the struck quark undergoes hard rescattering with a quark from the other nucleon, thus sharing the high momentum of the incoming photon. Due to the hard kernel of the quark-interchange interaction,  hard rescattering is expressed through the helicity amplitudes of high-momentum-transfer nucleon-nucleon ($NN$) scattering. 
The number of diagrams accounting for all possible quark interchanges between the outgoing nucleons is very large. However, the HRM allows to effectively account for this sum based on the observation that in the sum of all possible diagrams, the kernel of the hard rescattering can be identified with the quark-interchange kernel of the hard elastic $NN$ scattering. The latter allows to substitute the sum of the incalculable part of the break-up amplitude with the helicity amplitudes of hard elastic $NN$ scattering.

Given the $NN$ helicity amplitudes, the HRM allows to calculate the amplitude of $\gamma d\rightarrow p n$ scattering without any free parameters. One important aspect of the model is that while the invariant energy 
that enters in the $NN$ amplitude is the same as the energy of the $\gamma d$  and final $pn$ systems,
\begin{equation}
s_{N} = s_{\gamma d} = s = M_d^2  + 2M_dE_\gamma, 
\end{equation}
the  invariant momentum transfer $t_N$ that enters in the $NN$ amplitude is less than the one corresponding to 
the   $\gamma d\rightarrow pn$  reaction, $t = (k_\gamma - p_{1f})^2$:
\begin{equation}
t_{N}  \approx (k_\gamma + {p_{d}\over 2} - p_{1f})^2  = {t\over 2} + {m_N^2\over 2} - {M_d^2\over 4},
\label{t_tn}
\end{equation}
where  $m_N$ and $M_d$ are the nucleon and deuteron masses, $E_\gamma$ is the photon energy in the lab frame, and $k_\gamma$,  $p_d$, and $p_{1f}$ are the four-momenta of the incoming photon, deuteron, and outgoing 
proton, respectively ($u_{N}  \approx (k_\gamma + {p_{d}\over 2} - p_{2f})^2$).  This stems from the fact that the HRM model corresponds to double scattering in which each scattering process carries the half of the total $t$. 
If we introduce the center-of-mass scattering angle for  the $\gamma d\rightarrow pn$ reaction, 
$\theta^{\gamma d}_{c.m.}$, and the similar angle for 
$pn\rightarrow pn$ scattering, $\theta^{pn}_{c.m.}$, one obtains from Eq.(\ref{t_tn}):
\begin{eqnarray}
cos(\theta_{c.m.}^{pn}) &= &1 - {(s-M_d^2)\over 2(s-4m_N^2)}
{(\sqrt{s}-\sqrt{s-4 m_N^2}\cos(\theta^{\gamma d}_{c.m.}))\over \sqrt{s}}\nonumber \\  &&+ {4m_N^2 - M_d^2\over 2(s-4m_N^2)}.
\label{theta_cmn}
\end{eqnarray} 
This relation demonstrates that the $pn$ rescattering amplitudes enter at smaller angles than 
$\theta_{c.m.}^{\gamma d}$.  For example, $\theta_{c.m.}^{\gamma d}  = 90^\circ$  corresponds to 
 $\theta_{c.m.}^{pn} \approx 60^\circ$  in the elastic $pn$ amplitude.

Using the available experimental values of $pn$ scattering amplitudes, the 
HRM prediction for the differential cross section, as well as for the angular distribution of the hard $\gamma d \rightarrow pn$ 
reaction, resulted in a reasonable agreement with the data~\cite{Deute24, Deute09}.  The very same approach also 
allowed to successfully predict  the cross-section behavior of the hard break-up of the $pp$ pair in the  $\gamma+^3$$He\rightarrow pp + n(slow)$ 
reaction~\cite{RNA-rev,misakgranados, pomerantz}.

\begin{figure}[ht]
\hspace{-1.0cm}
\centering\includegraphics[scale=0.4]{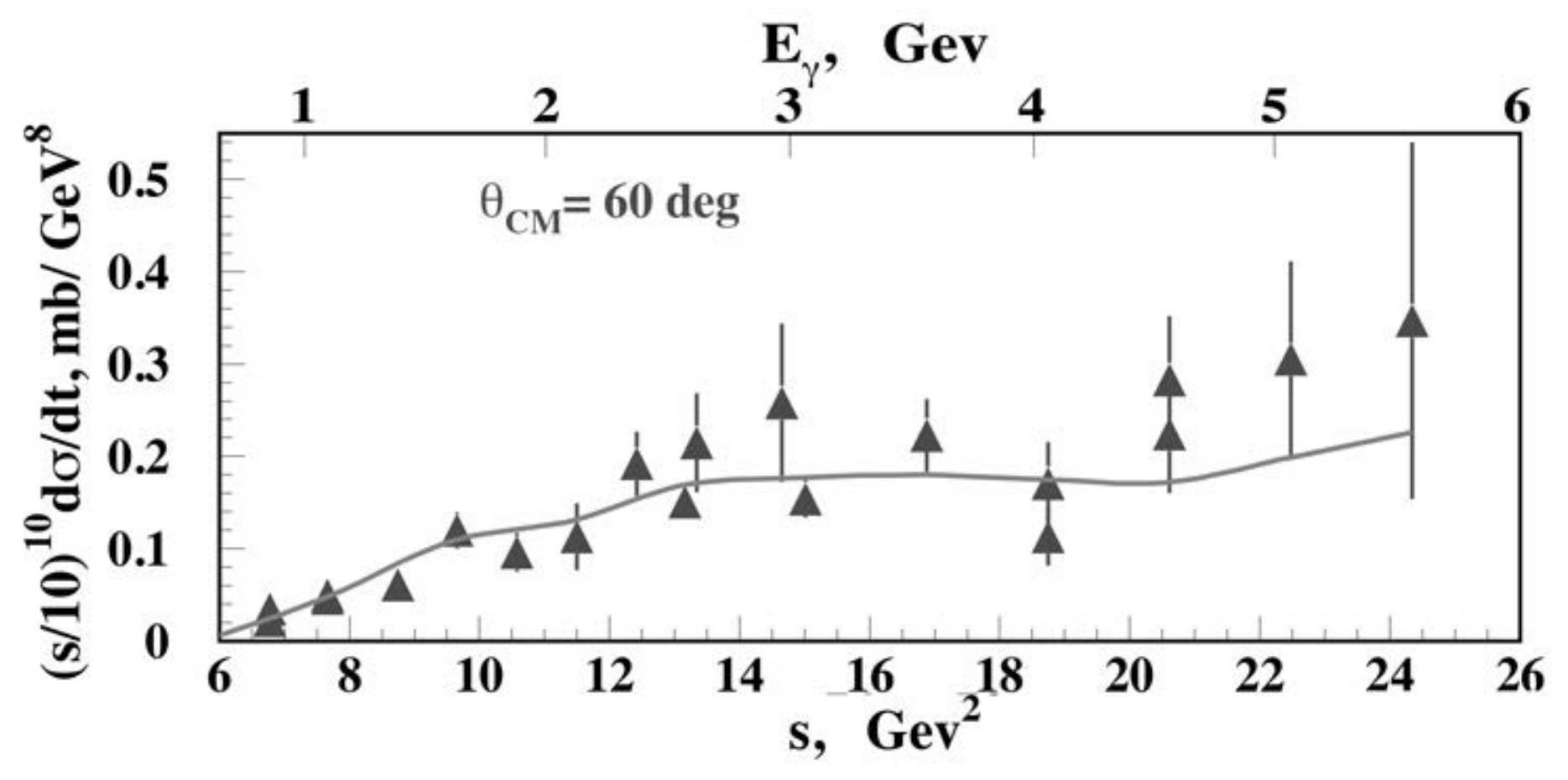}
\caption{ The $s^{10}$ scaled differential cross section of elastic $pn$ scattering as a function of $s$ and $E_\gamma$. Data are from Refs.~\cite{dataforAmplitude1, dataforAmplitude2} and the curves correspond to the HRM model fits used to calculate predictions for $\Sigma$.}
\label{pn60}
\end{figure}

Recently, calculations of the beam-spin asymmetry within HRM have been  updated using the progress made in describing 
$pn$ helicity amplitudes~\cite{misakgranados, misakgranados2}.   This progress is based on the recent observation of the symmetry structure of valence-quark wave function of the nucleon as well as the new parametrization of the $pn$ 
amplitude in the large $\theta_{c.m.}$ region. The new parametrization of helicity amplitudes of elastic $pn$ scattering is based on the di-quark model of the valence-quark wave function of the nucleon in which the relative phases between scalar and vector di-quarks are fixed. Figure~\ref{pn60} shows the $s^{10}$-scaled $p n \rightarrow pn$ differential cross section as a function of invariant variable $s$ (a) and $E_\gamma$ (b) at $\theta_{c.m.}^{pn}=60^\circ$ used to describe the energy dependence of the $\theta_{c.m.}^{\gamma d}=90^\circ$ helicity amplitudes. 
Using these fits, predictions were made for the beam-spin asymmetry at $\theta_{c.m.}^{\gamma d}=90^\circ$ (dotted-dashed line in Fig.~\ref{NewResultsFig2aa}). 

Due to the relation of Eq.~(\ref{t_tn}), angles other than $\theta_{c.m.}^{\gamma d}=90^\circ$ correspond to much smaller values of $t_N$ or $u_N$ and as a result, for photon energies between 1.1 -- 2.3 GeV, $t_N$ or $u_N$ are too soft for the  HRM to be valid.

\subsubsection{Quark Gluon String Model (QGSM)}
Another approach to the problem of non-perturbative parton dynamics is used in the QGSM proposed
by Kaidalov~\cite{kaidalov, kaidalova}. Spin variables have been included into the QGSM in Refs.~\cite{QGSMspin1,QGSMspin2}.
This model describes the reaction through the exchange of three valence quarks with an arbitrary number of gluon exchanges. The exchanged nucleon is replaced by a nucleon Regge trajectory that represents the sum of the exchanged resonances. A nonlinear Regge trajectory provides the best description of the data.
 In a general sense, the QGSM is a microscopic (nonperturbative) model of Regge phenomenology
for the analysis of exclusive and inclusive hadron-hadron and photon-hadron reactions
at the quark level. Originally, the QGSM was formulated
for the case of small scattering angles ({\it i.e.}, low momentum transfers).
Later, Kondratyuk {\it et al.} extrapolated the QGSM amplitudes to the case of
large-angle deuteron photodisintegration~\cite{ Deute22, Deute23, Deute23}.
The model fixes all but two of its free parameters from other processes, and fixes the remaining two using the experimental data on the deuteron photodisintegration cross section. It provides predictions for the angular distribution of the differential cross section and is sensitive to spin-dependent effects, making predictions for polarization observables.

\subsection{\emph{Experimental status of deuteron photodisintegration}}

The extensive studies of the differential cross sections~\cite{Deute09, Deute04, Deute01, Deute02, Deute03, Deute06a, Deute05,Deute06} have shown
that the different theoretical models describe the available cross-section data on the angular and energy dependence
 with about the
same degree of success.

Prior to the measurement presented here, there were only three sets of polarization data for
deuteron photodisintegration at energies above 1 GeV. The beam-spin
asymmetry, $\Sigma$, was measured at Yerevan~\cite{Deute11, Deute12}; the induced proton
polarization, $P_y$, and the polarization transfers, $C_{x'}$ and $C_{z'}$, were
measured at JLab~\cite{Deute07, PhysRevLett.98.182302}. On the theoretical side, two calculations of the
spin observables are available, within the QGSM~\cite{Deute38} and HRM~\cite{Deute37}
frameworks. The  prediction of the QGSM model for  the longitudinal polarization transfer
$C_{z'}$ is in good qualitative agreement with the measured data, but the model makes no
prediction for the transverse polarizations $P_y$ and $C_{x'}$ due to their
sensitivity to the relative phases of the helicity amplitudes~\cite{Deute38}.
In this respect, calculations of $C_{z'}$ are more stable because they do not
depend on these phases but only on the moduli squared of the helicity amplitudes. The HRM model predictions, as determined by a parametrization of the $pn$ helicity amplitudes, are in qualitative agreement with the available data for  both $C_{x'}$ and $C_{z'}$. 

For the beam-spin asymmetry, $\Sigma$, there are only the Yerevan data in the energy range 0.8 -- 1.6 GeV and at
$\theta_{c.m.}=90^{\circ}$~\cite{Deute11, Deute12} (see Fig.~\ref{NewResultsFig2aa}). Unfortunately, the data at $E_{\gamma}\approx 1.4$ -- 1.6 GeV
have large uncertainties and do not allow us to constrain the available models. 
Nevertheless,
the Yerevan data indicate that $\Sigma(90^{\circ})$ might be about $0.5$ at these
energies. In fact, the QGSM is able to accommodate a large 
beam-spin asymmetry of 0.5 at $E_{\gamma}\approx 1.6$ GeV and $\theta_{c.m.}=90^{\circ}$~\cite{Deute38}, while
the HRM is not able to do so~\cite{Deute37, Deute37a}.

The data  we present here on the beam-spin asymmetry, $\Sigma$, were obtained in an experiment that took place at JLab. Our results for $\Sigma$ cover photon energies between 1.1 and 2.3 GeV and nearly complete proton center-of-mass angles (between $\theta_{c.m.}=20^\circ$ and $\theta_{c.m.}=160^\circ$). A description of the experimental setup is given below.

\section{Experimental Setup}\label{ExpSet}
Deuteron photodisintegration was studied using the CEBAF Large Acceptance Spectrometer (CLAS)~\cite{RefExp2}, which was housed in Hall B at JLab.  
\begin{figure}[h!]
   \includegraphics[width=3.3in]{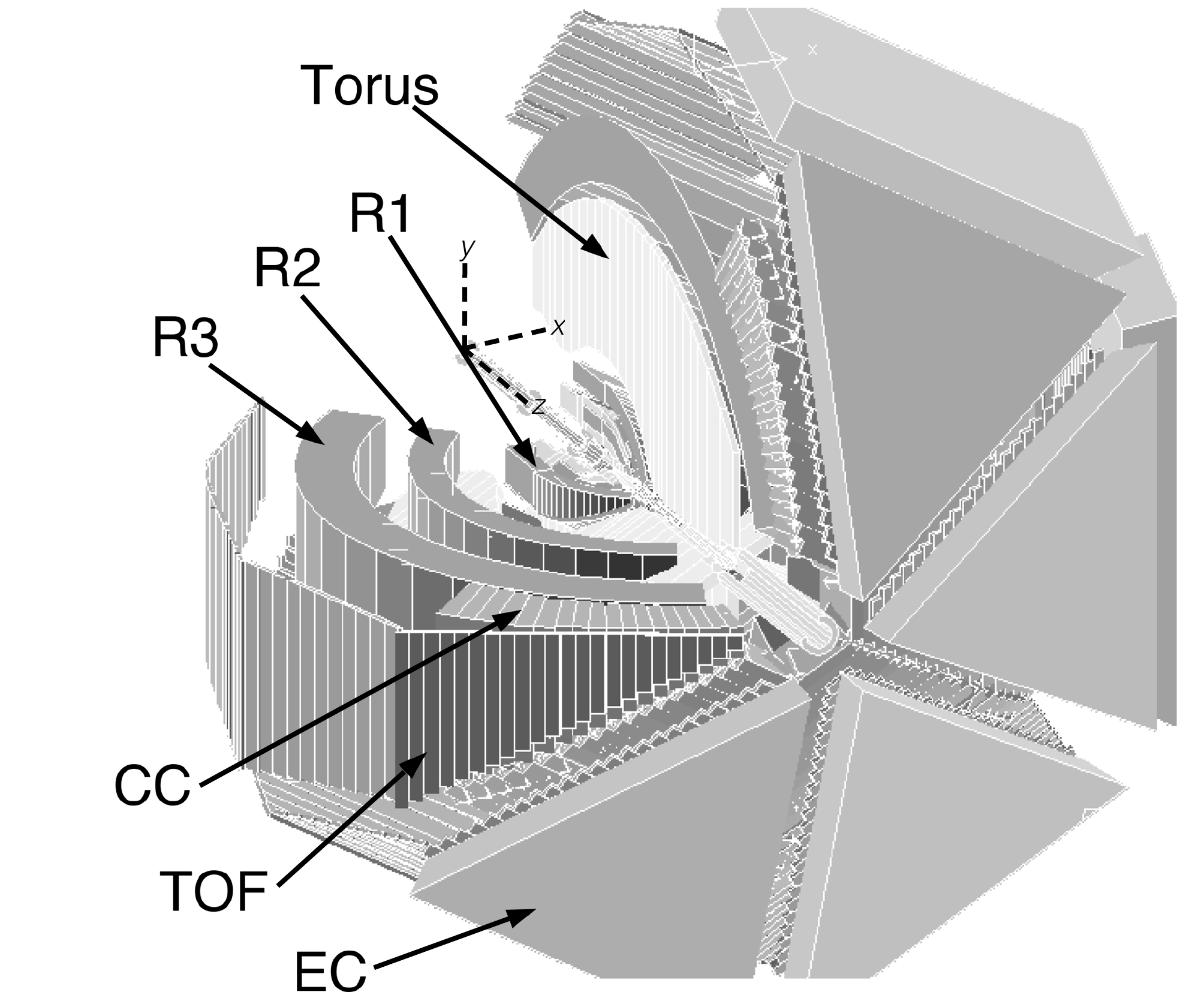} 
   \caption{A three-dimensional view of CLAS showing the torus magnet, the three regions of drift chambers (R1--R3), the \v{C}erenkov counters (CC), the time-of-flight detector (TOF), and the electromagnetic calorimeters (EC). The CLAS reference frame, also indicated in the here, was defined with the $z$ axis along the beam line and the $y$ axis perpendicular to the horizontal. Figure taken from~\cite{RefExp2}. }      \label{FigtheCLAS}
\end{figure} 
CLAS provided efficient detection of particles over a large solid angle. Six superconducting coils produced a non-uniform toroidal magnetic field and divided CLAS into six identical magnetic spectrometers (sectors) as shown in Fig.~\ref{FigtheCLAS}. 
Each sector contained three regions of drift chambers (Region 1-- R1, Region 2 --  R2, and Region 3 -- R3) that were used to track charged particles and reconstruct their momenta~\cite{RefExp11}, scintillator counters (TOF) for particle identification based on time of flight~\cite{RefExp15}, \v{C}erenkov counters (CC) to identify electrons (not used in this experiment)~\cite{RefExp14}, and electromagnetic calorimeters (EC) to identify electrons and neutral particles~\cite{RefExp16}.  

The geometry of CLAS allowed particle identification and momentum determination in a large portion of the full solid angle. Charged particles with laboratory polar angles between $8^\circ-140^\circ$ (this range varies depending on the target length and position) were tracked over approximately 83\% of the azimuthal angle with  1-mrad polar and 4-mrad azimuthal angular resolutions. A current of $-1500~$A in the torus magnet produced a magnetic field that bent negatively-charged particles away from the beamline. The charged-particle tracking system provided momentum resolution of about 0.5\%. Real-photon experiments made use of a start counter (ST), which was composed of 24 scintillator paddles that surrounded the target~\cite{RefExp16a}. The start counter was used in the event trigger and to determine the time at which nuclear reactions occurred in the target. 

A linearly polarized real-photon beam was produced via coherent bremsstrahlung using a 50-$\mu$m thick diamond radiator, which was positioned on a goniometer. The photon beam was then strongly collimated to enhance the linear polarization. The characteristics of the photon energy spectrum, such as the position of the coherent peak, and the degree of photon polarization, were controlled by the incident electron energy and the orientation of the crystal radiator with respect to the beam~\cite{KenLivingCLAS2011}. Electrons that produced bremsstrahlung photons were analyzed in the Hall-B tagging spectrometer (tagger)~\cite{RefExp10}, which consisted of a dipole magnet and scintillator hodoscopes. The tagger allowed the determination of the incident photon energy by identifying the hit position of the scattered electron in the hodoscope plane. It provided a tagging range between 20\% and 95\% of the incident electron-beam energy. The size of the scintillator paddles varied such that an energy resolution of about 0.1\% of the incident electron-beam energy was achieved.
The time of the scattered electron in the hodoscope plane was also measured with a resolution of better than 150~ps and was used to identify the photon that initiated the event detected in CLAS~\cite{RefExp10}. 

The  target used in this experiment was a 40-cm-long, conically shaped cell, with a radius of 2 cm at its widest point, filled with liquid deuterium. The target cell was placed such that its downstream end cap was at the center of CLAS.

\section{Event Selection and Reaction Reconstruction}\label{analysis}
The data used for this study were obtained during the CLAS g13b data-taking period, which was part of the E-06-103 experiment~\cite{g13exp} and took place from mid-March through June 2007. During this period about 30 billion triggers were recorded using a linearly-polarized photon beam. The photon-polarization vector was rotated between two orthogonal directions: parallel and perpendicular to the horizontal detector mid-plane, referred to as {\it Para} and {\it Perp}, respectively. Data for six nominal coherent-edge positions, 200-MeV apart between 1.3 and 2.3 GeV, were collected. These data were collected using 8 different incident electron-beam energies as shown in Table~\ref{table1}.
\begin{table}[htdp]
\begin{center}
\begin{tabular}{l|c}\hline\hline
$E_\gamma$ (GeV) & $E_e$ (GeV) \\\hline
1.3 & 3.302, 3.914, 4.192 \\
1.5 & 4.065, 4.475 \\
1.7 & 4.065, 4.748 \\
1.9 & 5.057 \\
2.1 & 5.057, 5.157 \\
2.3 & 5.157\\\hline
\end{tabular} 
\end{center}
\caption{Different electron beam energy settings used for the six nominal coherent-edge positions during g13b.}\label{AnaIIPhotonSett}\label{table1}
 \end{table}

 The trigger during g13b was relatively loose, a single-charged-particle trigger, which led to accumulation of data for a number of photoproduction reactions. In this study, all events with only one positively-charged track were analyzed based on the missing-mass technique.  Below we give a detailed description of the procedure followed to reconstruct the reaction  $\gamma d\rightarrow p n$.

\subsection{Proton identification}\label{PID}
\begin{figure}[!ht]
   \includegraphics[width=3.5in]{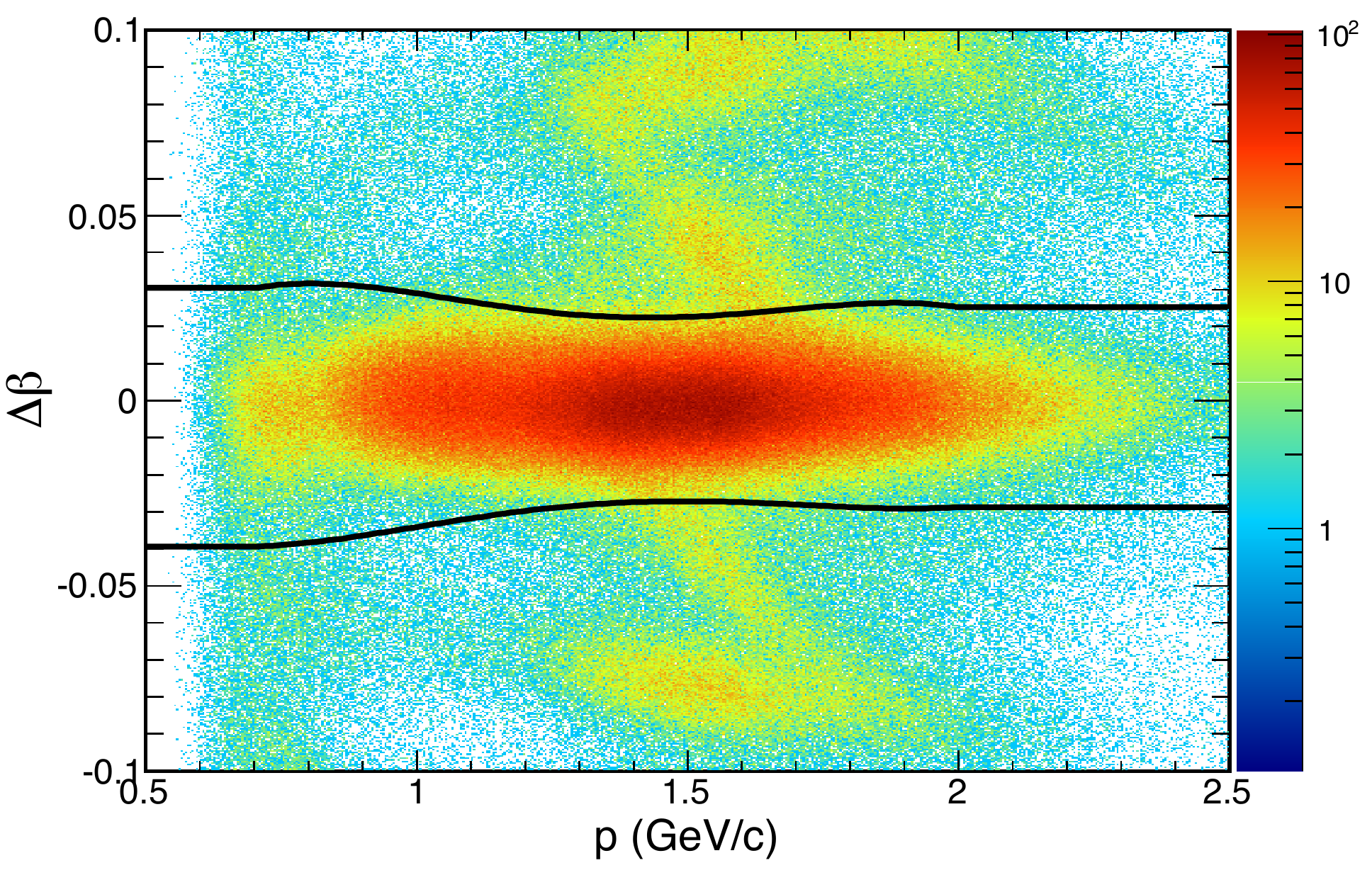} 
   \caption{(Color online) $\Delta\beta$ as a function of $p$. The lines show the $\pm3\sigma$ cut from the mean applied to identify protons.  }      \label{Figpid}
\end{figure}

Proton identification was done by comparing two independent estimates of the detected particle's speed (in units of the speed of light, $c$): one, $\beta_{meas}$, obtained as the ratio of the measured path length from the vertex to the TOF and the measured time of flight, and the other obtained from the measured momentum and an assumption about the particle's mass ($m_{nom}$). The difference between the two independent estimates was constructed as
\begin{equation}
\Delta\beta=\beta_{meas}-\sqrt{\frac{p^2}{m_{nom}^2c^2+p^2}}.
\end{equation}
To identify the protons in our sample, $m_{nom}$ was set to be the nominal mass of the proton. Figure~\ref{Figpid} shows the event distribution of $\Delta\beta$ as a function of the particle's momentum, $p$. Proton events are clustered around $\Delta\beta=0$. The proton-identification procedure was refined by accounting for the dependence of $\Delta\beta$ on the momentum resolution of the detector. This was done by dividing the distribution shown in Fig.~\ref{Figpid} into 50-MeV/$c$-wide momentum bins, and then fitting the $\Delta\beta$ distribution for each momentum bin 
to a Gaussian to determine the mean, $\mu_{\Delta\beta}$, and the standard deviation, $\sigma_{\Delta\beta}$. The momentum dependence of $\mu_{\Delta\beta}\pm3\sigma_{\Delta\beta}$ was parametrized and used as a proton-identification cut. 
Fits were not performed for the ranges $p<0.7$~GeV/$c$ and $p>2.0~$GeV/$c$, due to poor statistics, and straight-line extrapolations were used as cuts. 
The proton identification cut is indicated by the black curves in Fig.~\ref{Figpid}.  
The diagonal bands in Fig.~\ref{Figpid} are formed by events that were assigned the wrong mass, $m_{nom}$ ({\it i.e.}, non-proton events), as well as accidental events that  were due to particles that did not originate in the same physics reaction as the trigger particle. 

\subsection{Photon selection}\label{PhotonSele}
\begin{figure}[!ht]
   \includegraphics[width=3.5in]{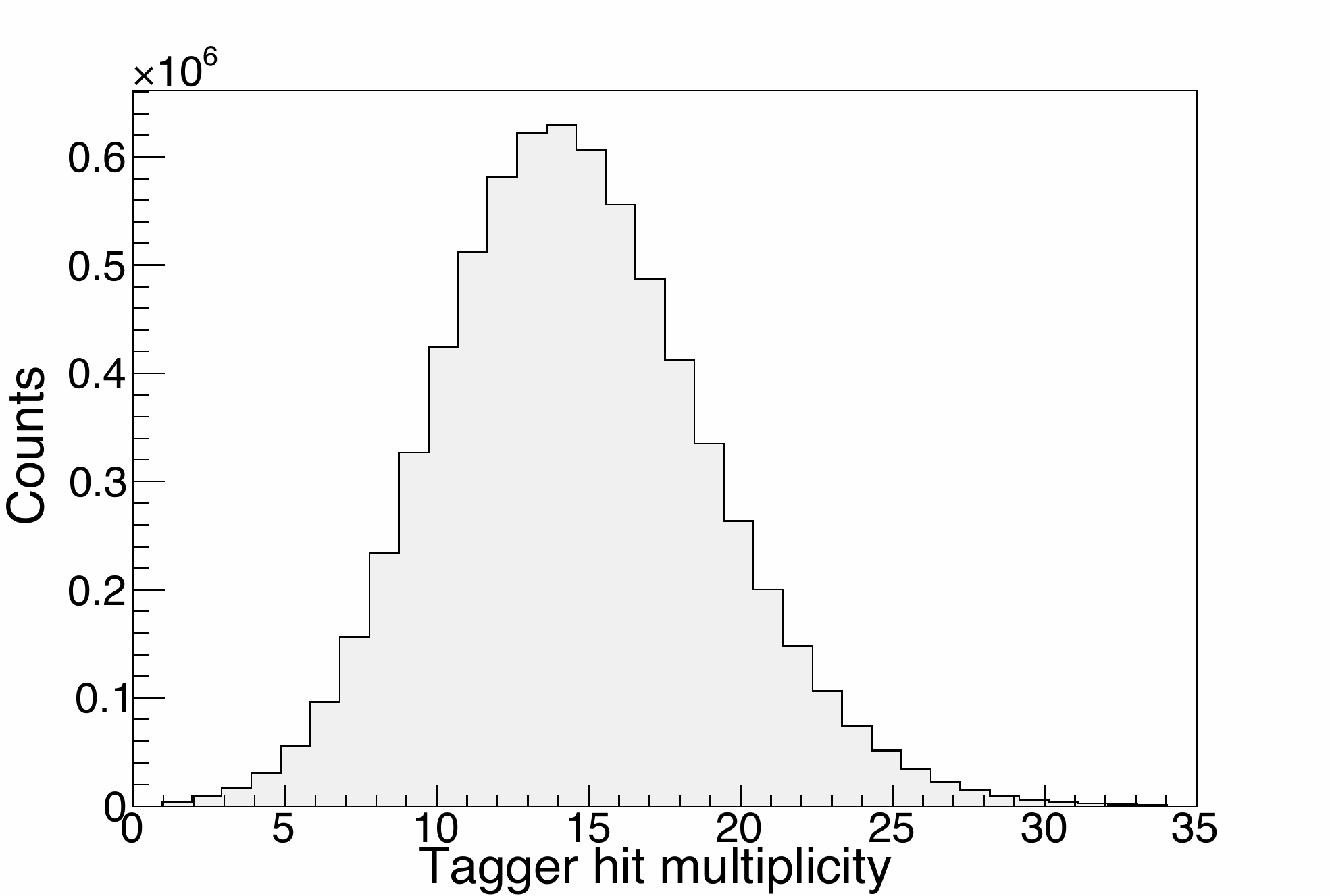} 
   \caption{Tagger hit multiplicity showing on average 14 photons as possible candidates for the true photon that initiated the event. }      \label{tagMult}
\end{figure}

\begin{figure}[!ht]
   \includegraphics[width=3.5in]{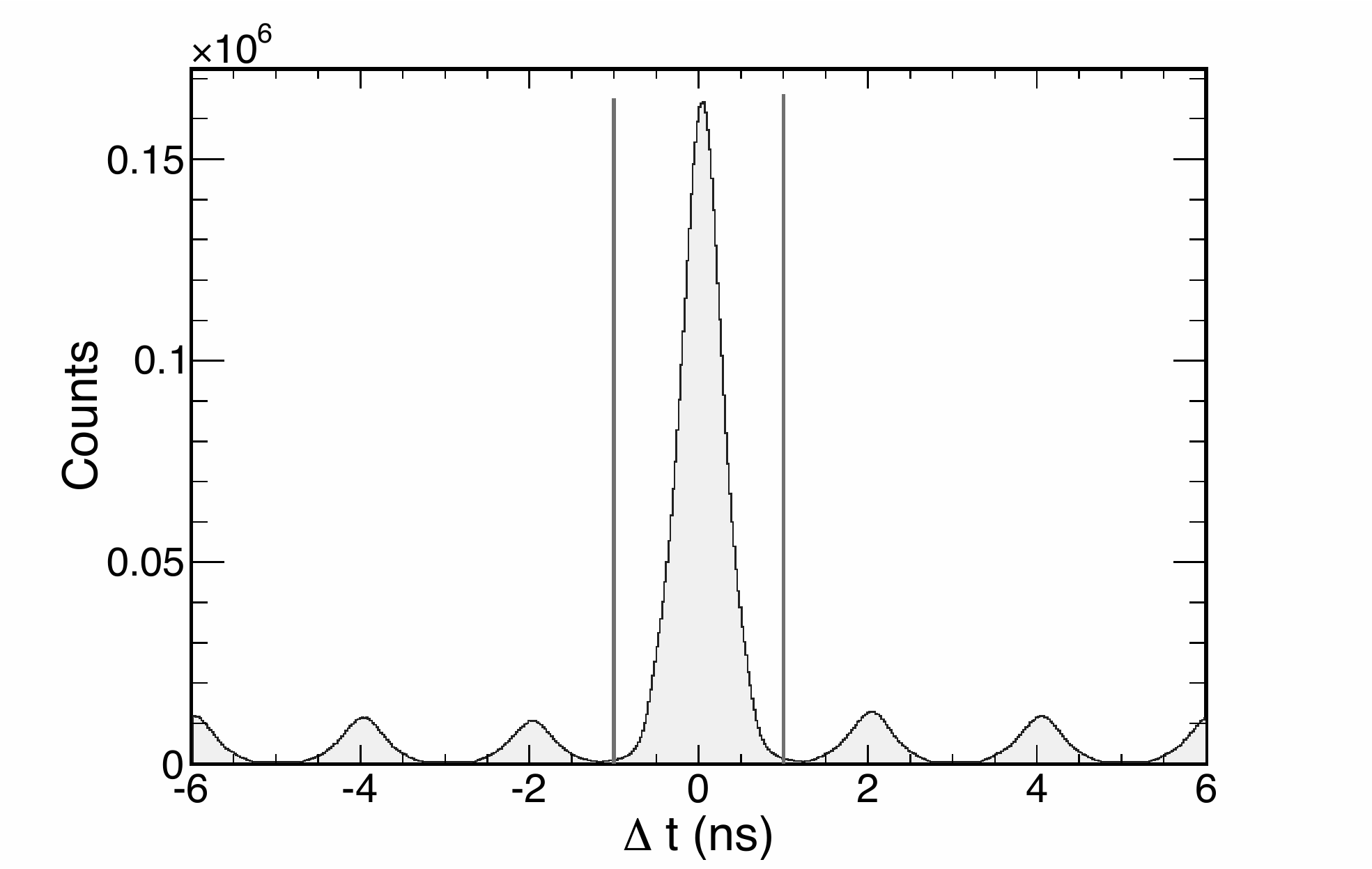} 
   \caption{ Coincidence time between all reconstructed photons and identified protons. The 2-ns bunch structure of the incident beam is evident. The solid vertical lines indicate the $\pm1$-ns cuts applied to identify the photon that initiated the reaction. }      \label{FigDeltT}
\end{figure}

During the g13b data-taking period 14 electron hits on average were recorded in the tagger for each trigger (see Fig.~\ref{tagMult}). Thus, for every event, there was a sample of $\sim14$ photons that could have produced the particle detected in CLAS.  
In order to identify the reaction of interest and to calculate kinematic variables, the photon that initiated the reaction must be selected from this sample. 
This was done by studying the time coincidence between the photon and the proton at the event vertex. 
The photon arrival time at the event vertex, $t_\gamma$, was calculated using electron timing information in the tagger hodoscope, whereas the proton vertex time, $t_v$, was calculated using timing information from CLAS. The coincidence-time distribution, $\Delta t=t_\gamma-t_v$, between all photon candidates and the identified proton is shown in Fig.~\ref{FigDeltT}. The peak centered at $\Delta t=0$ ns contains photon-proton coincident events. The small neighboring peaks at 2-ns intervals reflect the bunched nature of the incident electron beam. The photons in these neighboring peaks originate from other beam bunches (not the one that initiated the reaction) that came during the trigger window. The photon with a coincidence time within $\Delta t= \pm 1$~ns was selected as the photon that produced the proton. Events with two or more photons in this coincidence range were removed from further analysis. 
Overall, photons were unambiguously determined in about 78\% of all single-proton events. 

\subsection{Fiducial cuts}\label{FidCuts}
\begin{figure}[!ht]
   \includegraphics[width=3.5in]{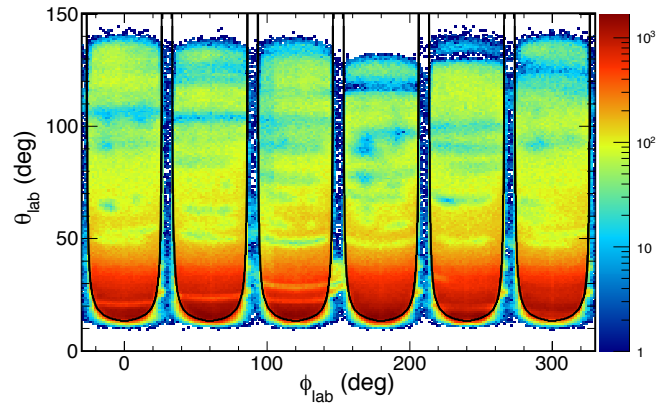} 
   \caption{(Color online) Angular distribution of protons as detected in the six sectors of CLAS. The black curves indicate the fiducial cuts applied to remove events that fell in regions where the acceptance changes rapidly.}      \label{figfiducial}
\end{figure}

Charged particles often escaped detection or failed track reconstruction in regions near the edges of the CLAS drift chambers. Typically, particles that hit the support frames or the cryostats of the torus magnet failed track reconstruction. In addition, the magnetic field close to the torus magnet varied rapidly with position and is not modeled very accurately. Therefore, particle tracks reconstructed in these regions are characterized by large systematic uncertainties, which in turn, propagate to a large systematic uncertainty in the reconstructed momentum. Furthermore, the method used to determine the beam-spin asymmetry assumes that the detector acceptance is constant within each kinematic bin, which is not true in the edge regions. To reduce systematic effects from these sources, we excluded events in which the detected particles fell in a region where the CLAS acceptance changes rapidly, by applying fiducial cuts.
To account for the shape of the drift chambers, these cuts were determined by studying the polar angle of reconstructed tracks as a function of the azimuthal angle. Figure~\ref{figfiducial} shows the angular distribution of protons in the six sectors of CLAS along with the applied fiducial cuts (black curves).
Additional cuts to remove inefficient regions within each sector did not affect our results since those inefficiencies were the same for all {\it Para} and {\it Perp} data, and thus canceled out in the ratio. 
 
\subsection{Event vertex cuts}\label{vertex}
\begin{figure}[!ht]
   \includegraphics[width=3.5in]{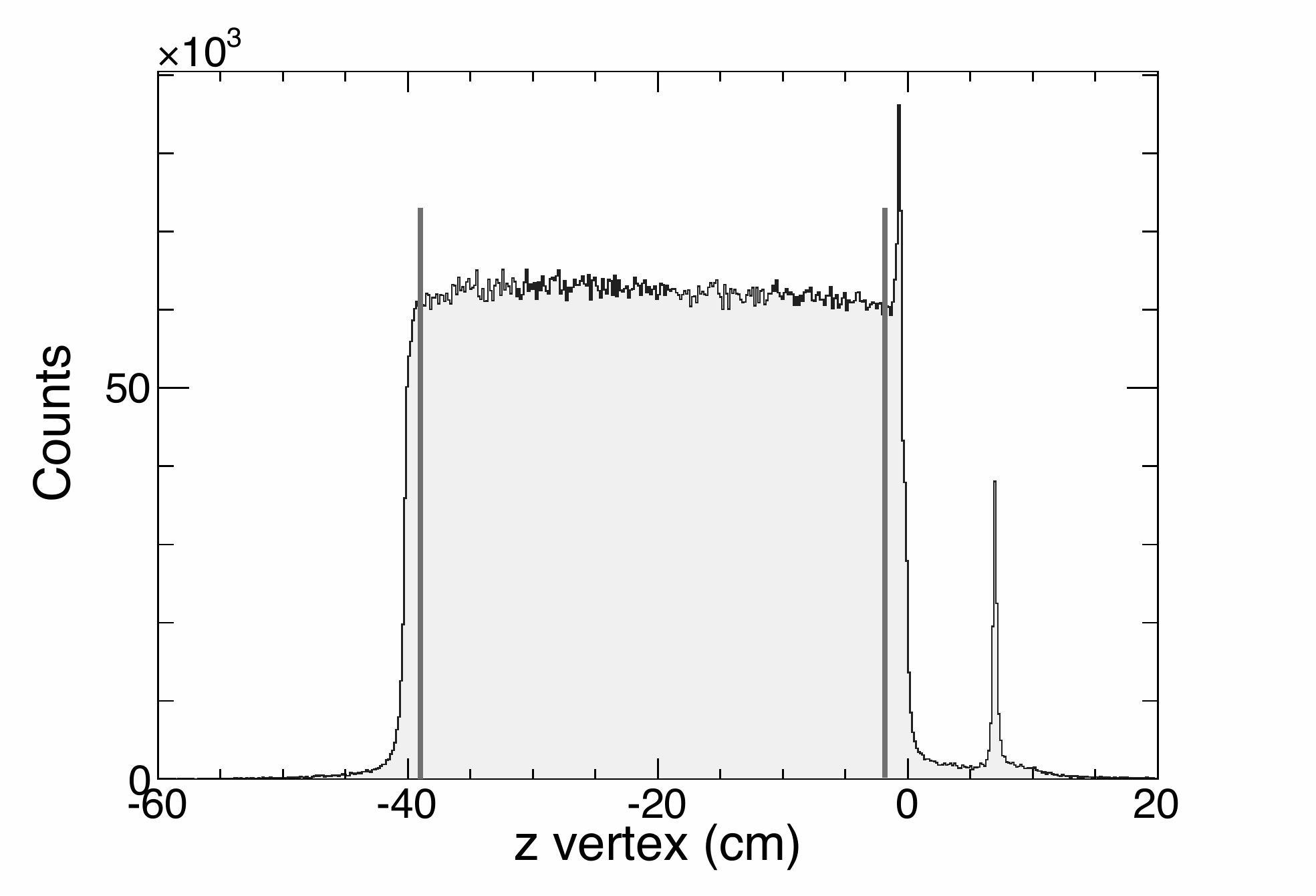} 
   \caption{$z$-component of the event vertex. The solid vertical lines indicate the cuts applied to select events that originated within the target. }      \label{zvert}
\end{figure}

In order to reduce background contribution due to events not originating in the target, we reconstructed and constrained the vertex of each event.
The event vertex was determined using the distance of closest approach between the proton track and the beamline position. The beamline position was determined for each data run using multi-charged-track events. Figure~\ref{zvert} shows the event distribution over the $z$-component of the vertex and the red lines indicate the cuts we applied to select events that originate in the target.
The contribution of events with $x$- and $y$-components of the event vertex outside the target (greater than $2~$cm) was negligible, and for this reason no cuts on the $x$ and $y$ vertex components were applied.

\subsection{Kinematic reconstruction and yield extraction}\label{Corrections}\label{missingmass}\label{backSub}
The magnitude of the momentum of a charged particle detected in CLAS was initially reconstructed under the assumption that the particle moved with constant speed throughout the detector. In order to obtain the momentum at the vertex, we corrected this initial estimate for the mean energy loss of the particle as it passed through the target, the start counter, and the air gap between the R1 drift chambers and the start counter~\cite{RefAnaI5}, as well as for the energy loss in the drift chambers, for drift chamber misalignments, and for small imperfections in the magnetic field map~\cite{RefAnaI6}. Corrections to the incident photon energies were also applied to account for a small gravitational sag in the tagger hodoscope~\cite{RefAnaI7}. 

\begin{figure}[h!]
   \includegraphics[width=3.5in]{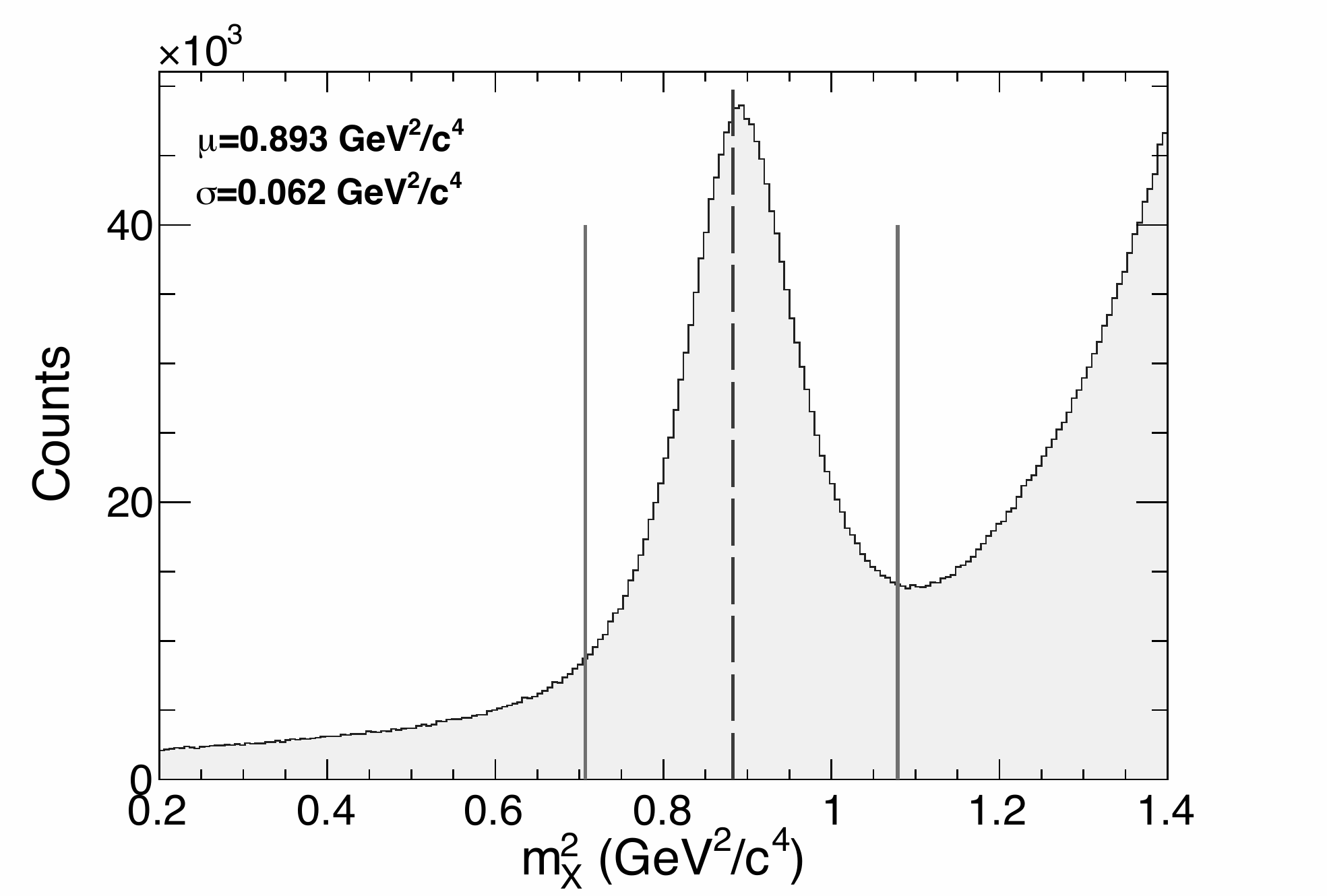} 
   \caption{Missing-mass-squared distribution of the reaction $\gamma d\rightarrow p X$. The peak corresponds to reactions where the missing particle is a neutron. The dashed vertical line indicates the nominal mass squared of the neutron, and the solid lines show the $\pm3\sigma$ missing-mass-squared cuts applied to selected deuteron photodisintegration events.}     
       \label{missingmassdistr}
\end{figure}

Deuteron photodisintegration events were then identified using the missing-mass technique. 
For each event, we calculated the missing mass squared, $m_X^2$, in the reaction $\gamma d\rightarrow p X$ using four-momentum conservation. In this calculation, the deuteron was considered to be at rest, the photon was identified as described in Sec.~\ref{PhotonSele}, and momentum and energy corrections were applied as discussed above. 
Figure~\ref{missingmassdistr} shows the missing-mass-squared distribution of all events that passed the selection cuts described in the previous subsections.  
Deuteron photodisintegration events are clustered in the peak centered at the nominal neutron mass squared and were selected by the application of a $\pm3\sigma$ cut on this distribution.

Figure~\ref{missingmassdistr} shows that in the $\pm 3\sigma$ missing-mass range of interest there was a non-negligible amount of background, in addition to deuteron photodisintegration events. This background contained primarily accidental events, and events from the reactions $\gamma d\rightarrow p p \pi^-$ and $\gamma d\rightarrow p n \pi^0$. 
The background varied from 5\% for the low photon-energy bins, to about 40\%  for the highest photon-energy bin.
The polarization observable of interest could be diluted or altered if the background was not removed from the deuteron photodisintegration sample.
To account for this background, a probabilistic weighting method was implemented. This method allowed for a signal-background separation on an event-by-event basis in a way that preserved all kinematic correlations~\cite{MwillProb1, RefAnaI8} by assigning each event  with a signal weight factor, $Q$, or equivalently, a background weight factor, $1-Q$. The $Q$-factors were then used to weight the contribution of each event in the ratio of polarized yields, $R(\phi)$ (see Sec.~\ref{Method}). A more traditional approach of fitting the missing-mass distribution in each kinematic bin with the same predetermined functions was also studied, and a comparison of the results was used to estimate the systematic uncertainty associated with the background subtraction method (see Sec.~\ref{SystUnc}). The probabilistic event weighting, amongst its many advantages, allows a more flexible kinematic binning without having to recalculate background contributions, and was thus the method of choice for background subtraction in this analysis.  

\begin{figure}[!]
   \includegraphics[width=3.5in]{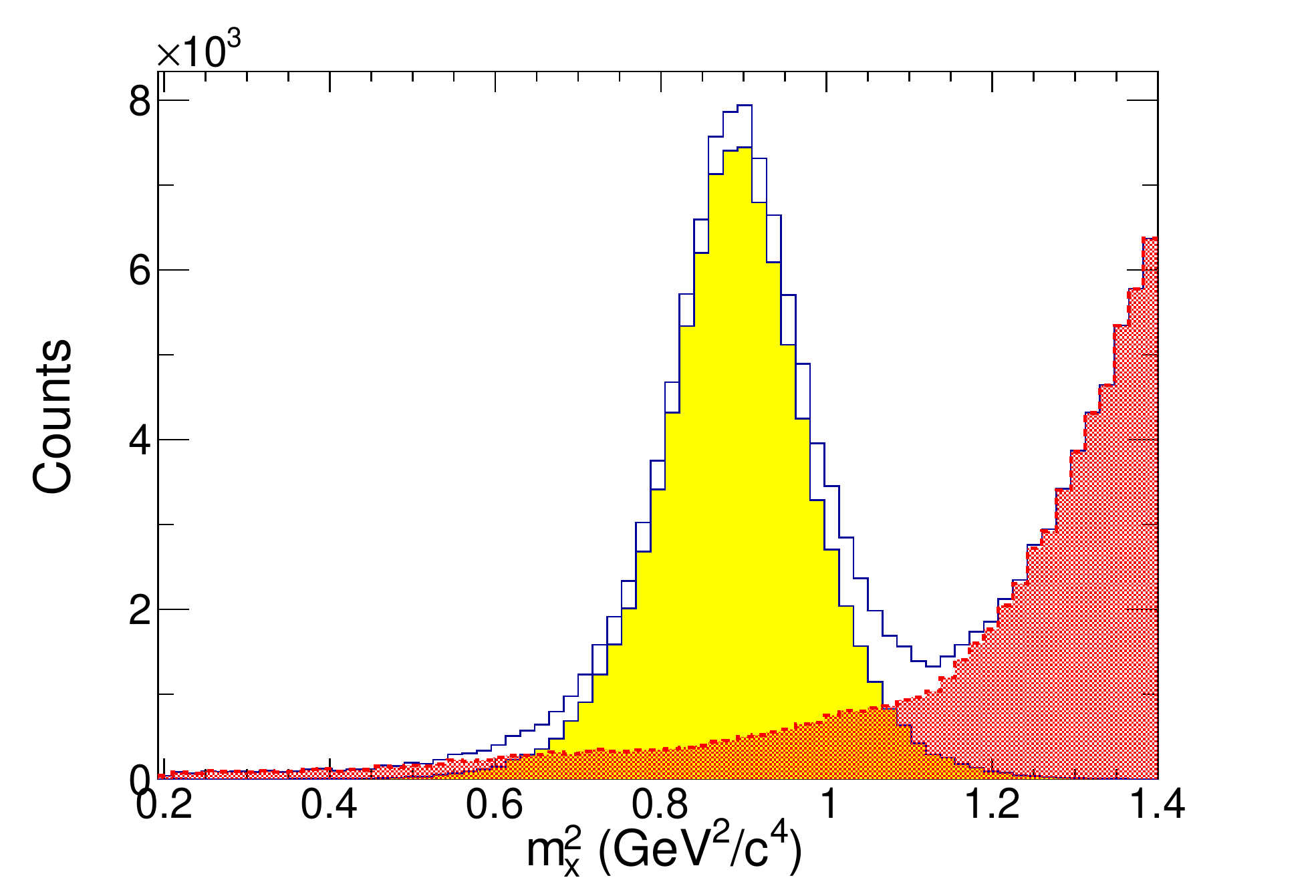} 
   \caption{(Color online) Missing-mass-squared distribution of the reaction $\gamma d\rightarrow p X$ of {\it Para} events in the 1.5--1.7 GeV photon-energy bin. The yellow histogram indicates background-subtracted deuteron photodisintegration events determined using the probabilistic event-weighting method~\cite{MwillProb1, RefAnaI8}. The red histogram indicates background events.  }      \label{Figmm}
\end{figure}

The signal-background kinematic correlations were preserved by dynamically binning data in photon energy $E_\gamma$, and proton angles $\theta_{c.m.}$ and $\phi$. The $Q$-factor of each event was determined by fitting the missing-mass distribution of the event's ``closest neighbors" with a predetermined function that described the signal and the background. The fitting was done independently on {\it Para} and {\it Perp} data, for each nominal coherent-edge position, and for each incident electron energy (see Tab.~\ref{table1}).  
The size of the dynamical bin used, which was defined by the number of ``closest neighbors", was determined by defining a metric in the proton angles $\theta_{c.m.}$ and $\phi$,
\begin{equation}
d_{ij}=\left[\frac{\cos\theta_i-\cos\theta_j}{2}\right]^2+\left[\frac{\phi_i-\phi_j}{2\pi}\right]^2. \label{distancebackground}
\end{equation}
For each event, $e_i$, in the data set, the distances  to all other events in the data set, $d_{ij}$ ($j=1,2\dots n$), were computed. Then, a predetermined number of events closest to $e_i$, $N_d$, was retained. These events were ``closest neighbors'' of $e_i$. In this analysis, 200 events were retained as closest neighbors. The missing-mass-squared distribution of the 200 closest neighbors was constructed and fitted with a Gaussian, describing deuteron photodisintegration events, and two exponentials, describing background events,
\begin{eqnarray}
g(m_X^2, A, \mu, \sigma)&=&A e^{-\frac{1}{2}\left(\frac{m_X^2-\mu}{\sigma}\right)^2},\\
b(m_X^2, A_1, A_2, B_1, B_2)&=&A_1 e^{A_2 m_X^2}+B_1 e^{B_2 m_X^2}.\;\;\;\;\label{backgroundfunctions}
\end{eqnarray}
These signal and background shapes were chosen since a fit using these shapes resulted in the best fit of the missing-mass-squared distribution. Other background shapes were studied ($1^{st}$ through $4^{th}$-order polynomials) without much success in adequately fitting the missing-mass-squared distributions. Systematic effects related to the choice of signal and background shapes were studied and are presented in Sec.~\ref{SystUnc}.
The fit parameters, $A$, $\mu$, $\sigma$, $A_1$, $A_2$, $B_1$, and $B_2$, determined from the fit, were used to determine the signal and background functions, $g(m_X^2)$ and $b(m_X^2)$, respectively. The missing-mass-squared value of event $e_i$, $m^2_{Xi}$, was used to calculate the strength of signal ($G_i$) and  background ($B_i$) for that event.
\begin{eqnarray}
G_i=g(m^2_{Xi})\nonumber\\
B_i=b(m^2_{Xi})\nonumber
\end{eqnarray}
The $Q$-factor of event $e_i$ was then calculated to be
\begin{equation}
Q_i=\frac{G_i}{G_i + B_i}.
\end{equation}
This procedure was repeated for each event, $e_i$, that passed the selection criteria discussed above and yielded a unique $Q$-factor, $Q_i$, for each $e_i$.
Different dynamic bin widths were studied with consistent results. A dynamic bin of 200 closest neighbors was chosen since it corresponds to a kinematic bin width similar to that used for the extraction of the beam-spin asymmetry. Specifically, for events in the lowest photon energy bin ($E_\gamma=1.1-1.3~$GeV), the dynamic bin size of 200 closest neighbors corresponded to about $5^\circ$ in polar and azimuthal angles for events produced at forward angles, and to about $10^\circ$ for events produced at backward angles. On the other hand, for events in the highest photon-energy bin ($E_\gamma=2.1-2.3~$GeV), the dynamic bin width of 200 closest neighbors corresponded to about $10^\circ$ in polar and azimuthal angles for events produced at forward angles, and to about $20^\circ$ for events produced at backward angles. 
Figure~\ref{Figmm} shows the result of the background subtraction method for {\it Para} events in the 1.5--1.7 GeV photon-energy bin.  Systematic uncertainties associated with the $Q$-factor determination were also taken into account and are discussed in Sec.~\ref{SystUnc}.

The yield of deuteron photodisintegration events $Y$ for each kinematic bin  was obtained as
\begin{equation}
Y=\sum_i^NQ_i,
\end{equation}
where $N$ is the number of events in the bin. The statistical uncertainty of the extracted yield in any kinematic bin is equal to the sum of the squares of the weights
\begin{equation}
\sigma^2_Y=\sum_i^NQ_i^2.
\end{equation} 

In summary, the background-subtraction method employed here successfully separated signal from background events while preserving all kinematic correlations. Different dynamic bin widths were studied with all of them yielding consistent results in the  observed quantities. The uncertainties associated with this method are well understood and taken into account.

\section{Photon Polarization}\label{PhotonPol}
\begin{figure}[!]
   \includegraphics[width=3.5in]{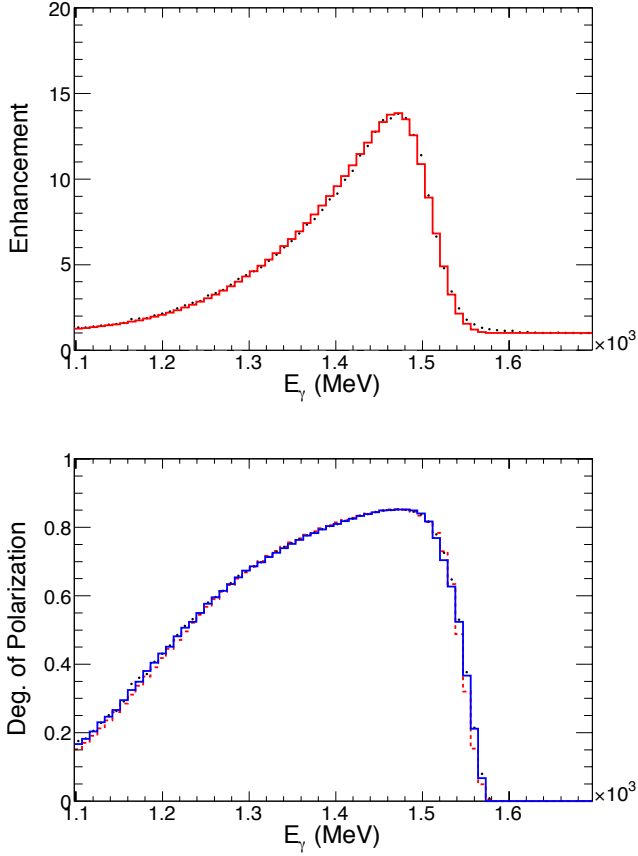} 
   \caption{(Color online) Enhancement distribution (upper plot) and calculated degree of photon polarization (lower plot) for events with the coherent-edge position at 1.5~GeV. The enhancement distribution is fitted with the analytic bremsstrahlung calculation~\cite{RefAnaII3b} and the degree of photon polarization is calculated (dashed red curve). The degree of photon polarization is then corrected for differences between the fit and the enhancement, as well as statistical fluctuations, using information from data with adjacent coherent-edge positions (blue curve). More information on the procedure is found in Refs.~\cite{KenLivingCLAS2011, nickthesis}. }      \label{FigPhotPol}
\end{figure}

For the determination of the beam-spin asymmetry, the degree of photon polarization had to be known. The latter was determined using an analytic bremsstrahlung calculation. The photon polarization was found by fitting the enhancement distributions with a theoretical calculation~\cite{RefAnaII3b} of the coherent spectrum. The enhancement distributions were obtained by dividing the coherent photon-energy spectrum by the photon-energy spectrum obtained from an amorphous radiator. The ratio, or enhancement distribution, rather than the coherent photon-energy spectrum, was constructed in order to remove systematic effects, such as counter-to-counter efficiency variations in the tagger, from the determination of the photon polarization.
 Parameters that are characteristic of the g13b data-taking period such as  electron-beam energy, beam collimation, beam divergence and angle, as well as fluctuations of the coherent radiator position and angle, the beam-spot size, and multiple scattering, were taken into account in the calculation.  Figure~\ref{FigPhotPol} shows the enhancement distribution for the coherent-edge position at 1.5~GeV fitted with an analytic bremsstrahlung calculation (upper plot). The lower plot shows the calculated photon polarization based on the enhancement fit (dashed red curve) and the corrected polarization that takes into account residual differences between the fit and the enhancement distribution (blue curve).  

The method and the procedure we used to determine the degree of photon polarization produce reliable results with small systematic uncertainties in regions where the enhancement is large.
Therefore, only events with photon energies above $E_{ce}-200$ MeV (where $E_{ce}$ is the coherent-edge position for the current event), 
were  kept for further analysis.  The average degree of photon polarization throughout the experiment was of the order of 75\%.
Details of the procedure to determine the photon polarization can be found in Refs.~\cite{KenLivingCLAS2011, nickthesis}.

\section{Determination of the Beam-Spin Asymmetry}\label{Method}
The beam-spin asymmetry, $\Sigma$, is related to the differential cross section of deuteron photodisintegration as:
\begin{equation}
\frac{d\sigma}{d\Omega}=\left(\frac{d\sigma}{d\Omega}\right)_{unpol}\left(1+P_{\gamma}\Sigma\cos[2\eta]\right).\label{AnaIIIEq1}
\end{equation}
This can be derived directly from the definitions of $\Sigma$ and the polarized cross section $\frac{d\sigma}{d\Omega}$, using the helicity amplitudes. 
In Eq.~(\ref{AnaIIIEq1}) $P_\gamma$ is the degree of linear polarization of the beam photon and $\eta$ is the azimuthal angle between the photon polarization vector and the reaction plane. In the case of {\it Para} events $\eta=\phi$ and in the case of {\it Perp} $\eta=\phi-90^\circ$, where the angle $\phi$ is the proton azimuthal angle measured in the CLAS reference frame. 
The determination of the beam-spin asymmetry is simplified by constructing the ratio of polarized yields, $R(\phi)=\left[Y(\phi)^{||}-Y(\phi)^{\perp}\right]/\left[Y(\phi)^{||}+Y(\phi)^{\perp}\right]$, where
{\small
 \begin{eqnarray}
Y(\phi)^{||,\perp}\sim& \int_{\phi-\frac{\Delta\phi}{2}}^{\phi+\frac{\Delta\phi}{2}}F^{||,\perp}\left(1\pm P_\gamma^{||,\perp}\Sigma\cos[2(\phi']\right)A(\phi')d\phi'\nonumber\label{ANAIIIeq8}\\
=&F^{||,\perp}\left(\Delta\phi\pm P_\gamma^{||,\perp}\Sigma\sin[\Delta\phi]\cos[2\phi]\right)A.\;\;\;\label{ANAIIIeq9}
\end{eqnarray}}
Equation~(\ref{ANAIIIeq9}) results from Eq.~(\ref{AnaIIIEq1}), with the parameter $F^{||,\perp}$ being the incident photon flux, $\Delta\phi$ the $\phi$-bin width used to bin the data, and $A$ the detector acceptance that is assumed to be constant within the $\phi$-bin (the effect of this assumption is investigated in Sec.~\ref{SystUnc}). The notations ${||}$ and $\perp$ indicate the orientation of the photon polarization vector.   

Substituting the expression in Eq.~({\ref{ANAIIIeq9}) in the definition of $R(\phi)$, the ratio becomes
{\small \begin{eqnarray}
R(\phi)&=&\frac{Y(\phi)^{||}-Y(\phi)^{\perp}}{Y(\phi)^{||}+Y(\phi)^{\perp}}\nonumber\\
&=&\frac{F_R-1+\frac{F_RP_R+1}{P_R+1}2\bar{P}\Sigma\frac{\sin[\Delta\phi]}{\Delta\phi}\cos[2(\phi-\phi_0)]}{F_R+1+\frac{F_RP_R-1}{P_R+1}2\bar{P}\Sigma\frac{\sin[\Delta\phi]}{\Delta\phi}\cos[2(\phi-\phi_0)]},\;\;\;\;  \label{ANA!!!eq10}
\end{eqnarray}}where $F_R=\frac{F^{||}}{F^{\perp}}$,  $P_R=\frac{P_\gamma^{||}}{P_\gamma^{\perp}}$, and $\bar{P}=\frac{P_\gamma^{||}+P_\gamma^{\perp}}{2}$. The parameter $\phi_0$ in Eq.~(\ref{ANA!!!eq10}) accounts for any systematic offset of the photon polarization vector from its nominal orientation.

Acceptance effects and any acceptance related systematic uncertainties, cancel out in $R(\phi)$. From the fit of the polarized-yield ratio, $R(\phi)$, to the function
\begin{equation}
F(\phi)=\frac{A-1+\frac{AB+1}{B+1}2C\cos[2(\phi-D)]}{A+1+\frac{AB-1}{B+1}2C\cos[2(\phi-D)]},\label{AnaIIIEq7}
\end{equation} 
 the following are determined:
\begin{itemize}
\item $A$: ratio of {\it Para} and {\it Perp} fluxes, $F_R$, 
\item $B$: ratio of {\it Para} and {\it Perp} polarizations, $P_R$,
\item  $C$: product of average polarization and asymmetry, $\bar{P}\Sigma\frac{\sin[\Delta\phi]}{\Delta\phi}$,
\item $D$: offset of the photon polarization vector, $\phi_0$.
\end{itemize}

The fitting was optimized by fixing three of the four parameters using independent methods. The optimization was extensively studied along with any associated systematic uncertainties~\cite{NickCLASNOTE}. Specifically, for each photon-energy bin, the parameter $A$ was determined from a fit to the azimuthal distribution of $R(\phi)$ integrated over all proton angles in the reaction of interest.
Since the incident photon flux was constant for a given photon-energy bin, a fit to the integrated kinematic bin ensures adequate statistics to precisely estimate the photon flux ratio $F_R$. The parameter $B$ was calculated using the degree of photon polarization on an event-by-event basis (as obtained from the procedure described in Sec.~\ref{PhotonPol}). Finally, the parameter $D$ was obtained from fits to the high-statistics single-pion reaction, $\vec{\gamma} d\rightarrow p_s p \pi^-$~\cite{RefIntro3}. 
 \begin{figure}[hb]
   \includegraphics[width=3.5in]{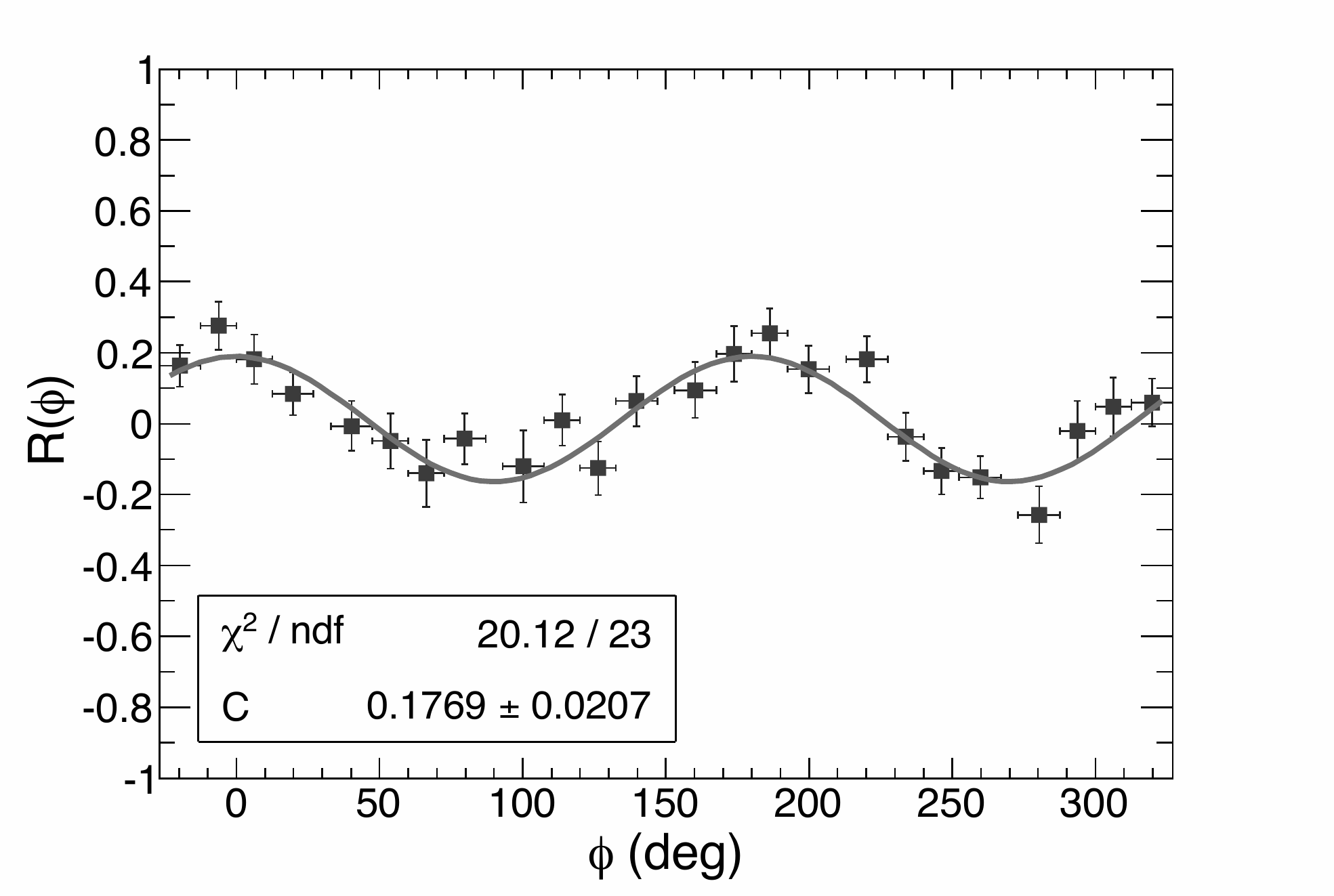} 
   \caption{  Ratio $R(\phi)$ for a specific kinematic bin ($E_\gamma=1.3-1.5$~GeV and $\theta_{c.m.}=120.0^\circ-132.5^\circ$) fitted to the function $F(\phi)$ (see text) to determine the beam-spin asymmetry for that bin.}      \label{FigRatioFit}
\end{figure}
The parameter $C$ was then determined by fitting the  ratio $R(\phi)$ for a bin in $\theta_{c.m.}$ and $E_\gamma$, fixing all other parameters, as shown in Fig.~\ref{FigRatioFit}. 

It is common in CLAS data analyses that the polarized-yield ratio is distributed in $\phi$ bins of variable width (smaller width in regions  in the middle of each CLAS sector and  larger width closer to the edges). This complicates the determination 
of the beam-spin asymmetry since the {\it correction factor}, $\frac{\Delta\phi}{\sin[\Delta\phi]}$, takes only a single value of $\Delta\phi$. One would intuitively expect that in the case of a fit to a variable-$\phi$-bin-width distribution, the correction factor would be some average over all $\phi$-bin widths; in fact, a good approximation of the correction factor can be calculated in this way ({\it i.e.}, $C.F.\approx\frac{\overline{\Delta\phi}}{\sin[\overline{\Delta\phi}]}$). To precisely quantify the value of the correction factor, we used Monte-Carlo data where we could control the true value of $\Sigma$.
 {\it Para} and {\it Perp} $\phi$ distributions were generated according to Eq.~(\ref{AnaIIIEq1}) and binned in the exact way CLAS data were binned, removing data that fell outside the CLAS fiducial regions. The function of Eq.~(\ref{AnaIIIEq7}) was fitted to the generated azimuthal distributions, and $\Sigma_{det}$ was obtained from the fit parameter $C$ as $\Sigma_{det}=C/P$. Then, the {\it correction factor}  for the variable $\phi$-bin widths was determined by fitting $\Sigma_{det}$ vs $\Sigma_{gen}$ with a first-order polynomial (see Fig.~\ref{corrfaca}). The slope of the fitted line gives the value of $C.F$. For the variable $\phi$-bin widths chosen for this analysis, the correction factor determined from this study is $C.F.=1.0094\pm 5\times 10^{-6}$.
 \begin{figure}[ht!]
   \includegraphics[width=3.5in]{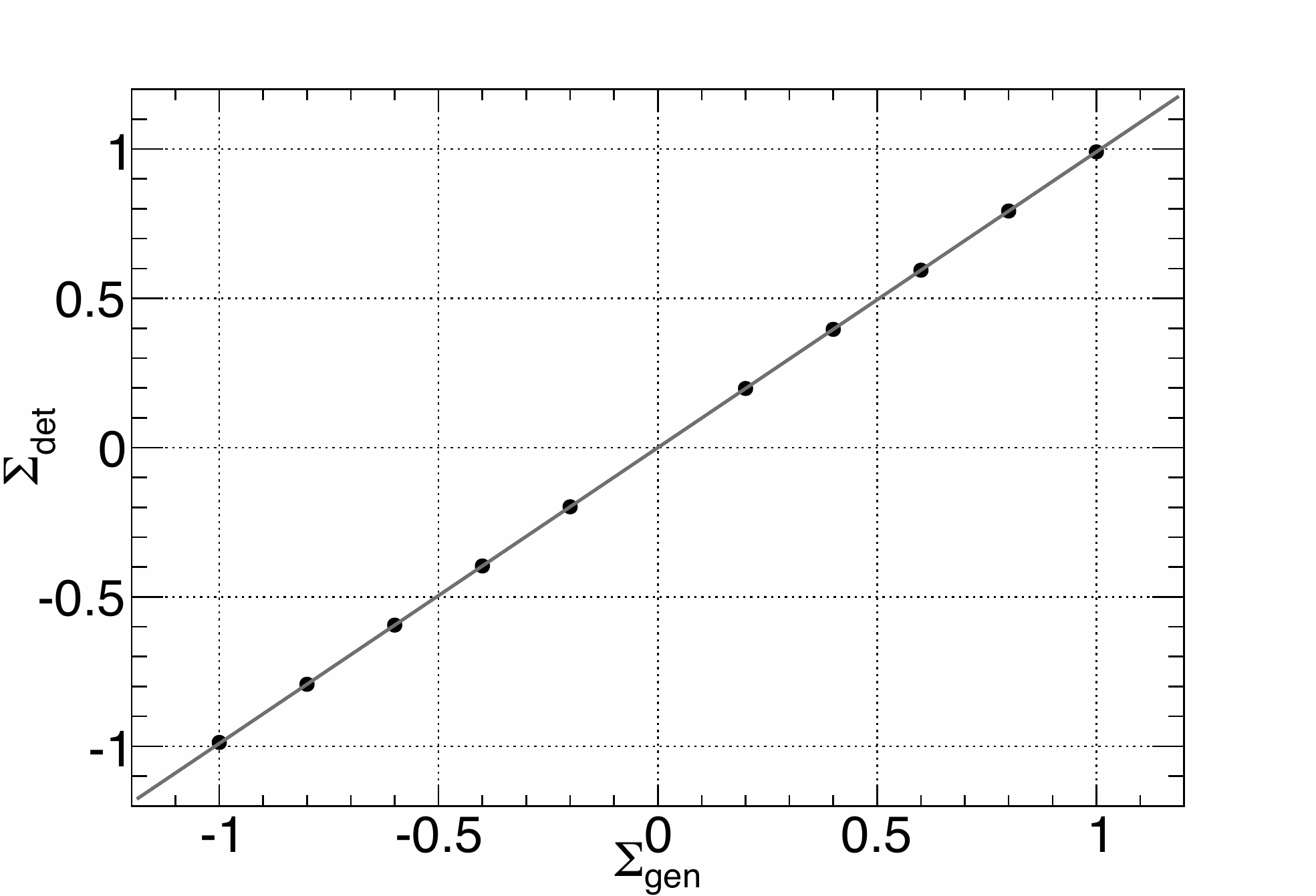} 
   \caption{Beam-spin asymmetries determined from fitting generated data as a function of the true value $\Sigma_{gen}$ for 36 variable $\phi$ bins (equal to an average bin width $\Delta\phi=13.50^\circ$). The correction factor due to a variable $\phi$-bin width, $C.F.$, is determined by fitting $\Sigma_{det}$ vs $\Sigma_{gen}$ to a first-order polynomial (see text for details). The statistical uncertainties of the points are smaller than the symbol size.}      \label{corrfaca}
\end{figure}
Further details can be found in Ref.~\cite{NickCLASNOTE}.

\section{Systematic Uncertainties}\label{SystUnc}
Systematic uncertainties due to various sources were studied and quantified. The uncertainties fall into several main categories: due to the $\phi$-bin method used to extract $\Sigma$, due to the choice of cuts for event selection, due to the choice of background shape, and due to the statistical uncertainties of quantities used to calculate $\Sigma$.
The parameters of these sources were varied within reasonable bounds and the sensitivity of the final result was checked against this variation. In several studies, generated data were used to determine the systematic uncertainties, and in others, the actual experimental data were used. 
A summary of the systematic uncertainties  is given in Table~\ref{TableSyst} indicating whether the source results in an absolute or relative uncertainty. 
\begin{table}[htdp]
\begin{center}
\begin{tabular}{l|l|c}\hline\hline
&\\
{\bf Source}&{\bf Estimate }&{\bf Type}\\\hline\hline
Variable $\phi$-bin width& $10^{-6}$ &1 \\
Detector acceptance& $<$1\%&2\\
Uncertainty of $\phi_0$ offset&  $10^{-6}$&1 \\
Uncertainty of $P_R$& 1\%&2\\
Uncertainty of $\bar{P}$&5\%&2\\
Uncertainty of $F_R$& $\sim$0.002&1 \\
Fiducial cuts& $\sim$0.014  &1\\
Particle ID cuts& $\sim$0.01  &1\\
Missing-mass cuts&$\sim$0.01 &1\\
Background subtraction method& $\sim$0.012&1 \\
Background shape& $\sim$0.01&1 \\
$Q$-factor&$\sim0.02$&1\\\hline\hline
\end{tabular} 
\end{center}
\caption{Systematic uncertainties contributing to the total systematic uncertainty of the beam-spin asymmetry. The estimated values are averaged over all kinematic bins. The type of the source indicates whether the uncertainty is absolute  (Type 1) or relative  (Type 2).  }\label{TableSyst}
 \end{table}
This section summarizes the studies performed to estimate the systematic uncertainties.
 
 \subsection{Variable $\phi$-bin width}
 The uncertainty of the {\it correction factor} that accounts for  the variable $\phi$-bin width has been propagated to the uncertainty of the beam-spin asymmetry using generated distributions. The value of this uncertainty was found to be of the order of $10^{-6}$ and is negligible compared to all other uncertainties~\cite{NickCLASNOTE}.

 \subsection{Detector acceptance}
Acceptance effects on the determined beam-spin asymmetry are twofold. Firstly, the $\phi$-bin method used to determine the beam-spin asymmetry relies on the assumption that the acceptance is constant within each $\phi$-bin, and thus can be taken out of the integral (see Eq.~(\ref{ANAIIIeq9})). Secondly, the detector acceptances of the {\it Para} and {\it Perp} data are assumed to be identical, and therefore cancel out in the ratio $R(\phi)$ (Eq.~(\ref{ANA!!!eq10})).

The effect of a non-constant acceptance within each $\phi$-bin has been investigated using simulated distributions. Details of the study can be found in Ref.~\cite{NickCLASNOTE}. The systematic uncertainty of $\Sigma$  due to the assumption of constant acceptance within each $\phi$-bin was found to be less than $1\%$. 

Data from the g13b data-taking period was collected in a way to minimize any difference between the acceptance  for the {\it Para} and {\it Perp} settings. This was accomplished by  changing the photon polarization between {\it Para} and {\it Perp} about every two hours.
Studies comparing the ratio of proton yields in adjacent TOF counters between {\it Para} and {\it Perp} data averaged over all runs within each  coherent-edge setting show that variations in the detector acceptance were within the statistical uncertainties. In addition, generated data using different acceptances for {\it Para} and {\it Perp} distributions of the size of the experimental variations showed negligible effects on the determined beam-spin asymmetry. The overall uncertainty of $\Sigma$ due to acceptance effects is thus less than 1\%.

\subsection{$\phi_0$ offset}
The systematic effect due to the uncertainty of the direction of the photon-polarization vector ($\phi_0$ offset) was  investigated using generated data~\cite{NickCLASNOTE}. The uncertainty of the  beam-spin asymmetry, which stems from the uncertainty of the $\phi_0$ offset,  was found to be of the order of $10^{-6}$  and is negligible compared to all other uncertainties.

 \subsection{Photon polarization}

 The effect of the uncertainty of the photon polarization  on the estimated value of $\Sigma$ is twofold. On one hand, the uncertainty of the photon polarization propagates into the uncertainty of the polarization ratio $P_R$, which is used as a fixed parameter in the fit (Eq.~(\ref{AnaIIIEq7})). This  affects the fit  and the uncertainty of the free-fit parameter $C$ that is used to determine $\Sigma$. On the other hand, the  uncertainty of the photon polarization propagates into the uncertainty of the average photon polarization  $\bar{P}$, which is used to calculate $\Sigma$ from $C$: $\Sigma=\frac{C}{\bar{P}}\frac{\Delta\phi}{\sin[\Delta\phi]}$. An independent study determined that the systematic uncertainty of the photon polarization was 7\%~\cite{DuggerCLASNOTE}. 
  Using the  value of $7\%$, the uncertainty of $P_R$ is calculated to be 
\begin{eqnarray}
 \Delta P_R&=&P_R\sqrt{\left(\frac{\Delta P_{||}}{P_{||}}\right)^2+\left(\frac{\Delta P_{\perp}}{P_{\perp}}\right)^2 }\nonumber\mbox{ and thus,}\\
 \Delta P_R&=& 0.1\times P_R.
 \end{eqnarray}
 The uncertainty of $\bar{P}$ is
\begin{eqnarray}
\Delta\bar{P}&=&\frac{1}{2}\sqrt{\left(\Delta P_{||}\right)^2+\left(\Delta P_{\perp}\right)^2}\nonumber\mbox{ and thus,}\\
\Delta\bar{P}&\sim& 0.05\times \bar{P},
\end{eqnarray} 
where we used $P^{\perp}\approx P^{||}\approx\bar{P}$.
The uncertainty of $\Sigma$ due to the uncertainty of $\bar{P}$ is then $5\%$.

To estimate the uncertainty of $\Sigma$ due to the uncertainty of $P_R$, generated distributions were produced and analyzed~\cite{NickCLASNOTE}. The study yielded an uncertainty of $\Sigma$ of less than 1\%.

In general, the uncertainties due to $P_R$ and $\bar{P}$ are highly correlated and should be treated together. However, it is evident that the uncertainty of the average polarization of $\sim5\%$ has a much bigger effect on $\Sigma$ than the uncertainty in the polarization ratio $P_R$ ($< 1\%$), and the former is quoted.

 \subsection{Incident photon flux}
The systematic uncertainty due to the uncertainty of $F_R$ was determined using studies similar to those for the $\phi_0$ offset~\cite{NickCLASNOTE}. From these the systematic uncertainty of the beam-spin asymmetry was found to be
\begin{equation}
\sigma^{sys}_{F_R} = 0.073 \cdot \sigma_{F_R}, 
\end{equation}
which on average corresponds to an uncertainty of 0.002.

  \subsection{Reaction selection cuts} 
The systematic uncertainty due to the choice of fiducial cuts was determined by varying the cuts from their nominal values to tighter values (the fiducial ranges of $\theta$ and $\phi$ were reduced by $\sim3^\circ$). The variation of the beam-spin asymmetry was found to be on average 0.014 and we report this value as the systematic uncertainty for this source.
 
Proton ID cuts were  varied between $2\sigma$ and $3\sigma$ using experimental data and the determined beam-spin asymmetries were compared. The variation of $\Sigma$ on average was 0.01 and we quote this value as the systematic uncertainty.

The missing-mass cuts, which select deuteron photodisintegration events, were varied between $2\sigma$ and $3\sigma$, and the determined beam-spin asymmetries were compared. This study accounts for possible leakage of events from background channels due to the non-Gaussian shape of the signal. On average the uncertainty of $\Sigma$ was  found to be 0.01.

 \subsection{Background subtraction}
The systematic uncertainty associated with the background subtraction is threefold. Specifically, there is a systematic effect associated with the predetermined shape of signal and background, which is used in the fits and the determination of  the $Q$-factor. In addition, there is an uncertainty associated with the background-subtraction method. Finally, there is an uncertainty associated with the $Q$-factors. This latter uncertainty can be determined by propagating the uncertainties of the fit parameters to $Q_i$. 

The uncertainty due to the assumption that the signal is Gaussian is accounted for in the systematic uncertainty associated with the missing-mass cut. The uncertainty due to the choice of the background shape was studied by comparing results obtained with a linear background and with the nominal background  (two exponentials -- see Eq.(\ref{backgroundfunctions})). This uncertainty was found to be of the order of $0.01$. 

The uncertainty associated with the background subtraction method itself was studied by comparing results from the probabilistic event-weighting method, which used a dynamic bin width, to results from a binned method in which the background subtraction was determined on a bin-by-bin basis. The study yielded a systematic uncertainty of 0.012.

The $Q$-factor weights have an uncertainty that depend on the dynamic bin width as well as on the goodness of the fit. 
Specifically,  the calculation of the uncertainty of $Q_i$ was done by propagating the uncertainties of the fit parameters, as
\begin{equation}
\sigma_{Q_i}=\sum_{jk}\frac{\partial Q_i}{\partial p_j} Cov(j,k)\frac{\partial Q_i}{\partial p_k}, \label{covprop}
\end{equation}   
where  $j$ and $k$ run over the number of the fit parameters, $p_i$ are the fit parameters, and $Cov(i,j)$ is the covariance matrix determined from the fit. The $Q$-factor uncertainties are highly correlated between events in the same kinematic bins due to the method of nearest neighbors.
Therefore, the  $Q$-factor uncertainties for each kinematic bin add up,
\begin{equation}
\sigma_{fit}=\sum_{i}\sigma_{Q_i},\label{addupdude}
\end{equation}   
where the sum is over the number of events in each $E_\gamma$, $\theta_{c.m.}$, and $\phi$ bin. Through several studies using the deuteron photodisintegration events, we were able to determine the systematic uncertainty of the beam-spin asymmetry, which is due to the uncertainty of the $Q$-factor value, as a function of the statistical uncertainty of $\Sigma$, $ \sigma_{stat}$, 
\begin{equation}
\sigma_\Sigma^{Q} =0.0093+0.176\cdot \sigma_{stat}.
\end{equation}
On average, this corresponds to an uncertainty of  0.02. 
More details on the determination of this uncertainty can be found in Ref.~\cite{NickCLASNOTE}.

\section{Results and Discussion}\label{Results}
Using the data from g13b, the beam-spin asymmetry was determined for incident photon energies from $E_\gamma=$1.1 to $E_\gamma=$2.3 GeV and proton angles between $\theta_{c.m.}=25^\circ$ and $\theta_{c.m.}=160^\circ$. In the following sections we report the energy and angular dependence of the beam-spin asymmetry and associated uncertainties, which were determined as outlined in the previous sections. 
\subsection{Energy distributions}\label{Edistributions}
Figure \ref{NewResultsFig2aa} shows our $\theta_{c.m.}=90^\circ$ data for $\Sigma$ compared to the available data from Yerevan and to the model predictions from QGSM and HRM. The precise CLAS data are in good agreement with the published Yerevan data and increase the kinematic coverage up to photon energies of 2.3 GeV. The linear energy dependence predicted by the QGSM is not confirmed by the data: the model predicts larger asymmetries than the data at lower photon energies and lower values than the data at higher photon energies. The updated HRM reproduces the general shape of the energy dependence. Especially, it describes the increase to higher asymmetries observed in the data between photon energies 1.6 and 2.0 GeV. In the HRM, this increase stems from features of the $pn$ scattering amplitude. However, the model underpredicts the values of the asymmetries over the entire energy range. 
\begin{figure*}[!]
   \begin{center}
   \includegraphics[ width=6.3 in]{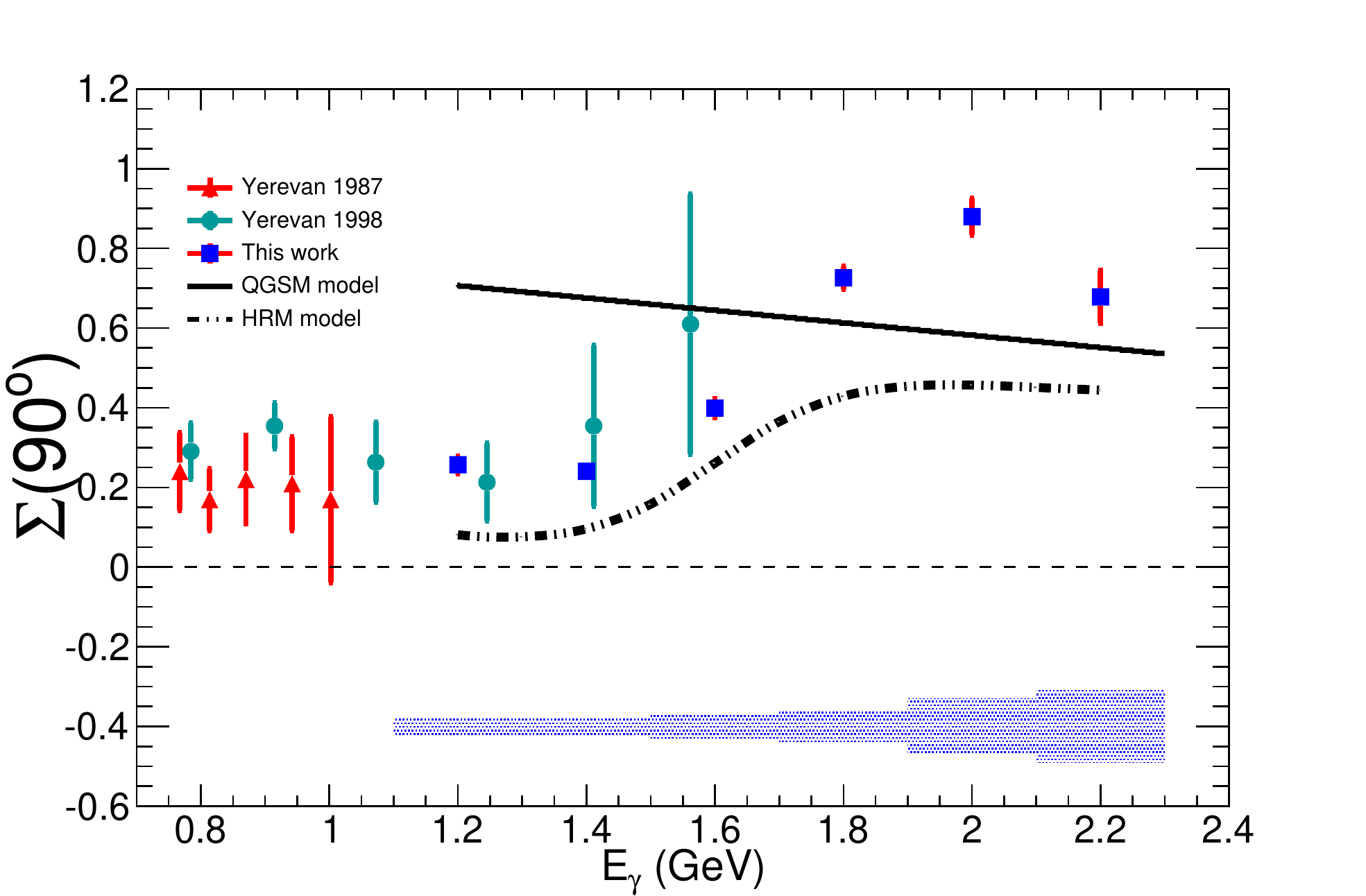} 
   \caption{(Color online) Beam-spin asymmetry as a function of the incident photon energy for $\theta_{c.m.} = 90^\circ$. The red (triangle) and cyan (circle) points show the results from Yerevan~\cite{Deute11, Deute12}, whereas the blue square points are the results from the present work. The solid and dash-dotted lines are the QGSM~\cite{Deute38} and HRM~\cite{Deute37, Deute37a} predictions, respectively. The blue band indicates the systematic uncertainties of the present CLAS measurements.}      \label{NewResultsFig2aa}
   \end{center}
\end{figure*}

Figure~\ref{NewResultsFig2a} shows the energy dependence of $\Sigma$ for four different proton center-of-mass angles. The width of the photon-energy bin is kept constant at 200 MeV, whereas the width of the angular bins varies in an attempt to have similar statistical uncertainties. The results indicate  positive asymmetries for angles larger than $50^\circ$ and negative asymmetries for forward-going protons. The $\theta_{c.m.}=90^\circ$ result displays the largest asymmetry and suggests a local maximum at $E_\gamma=2.0$~GeV.
\begin{figure*}[h!]
   \begin{center}
   \includegraphics[ width=3.0 in]{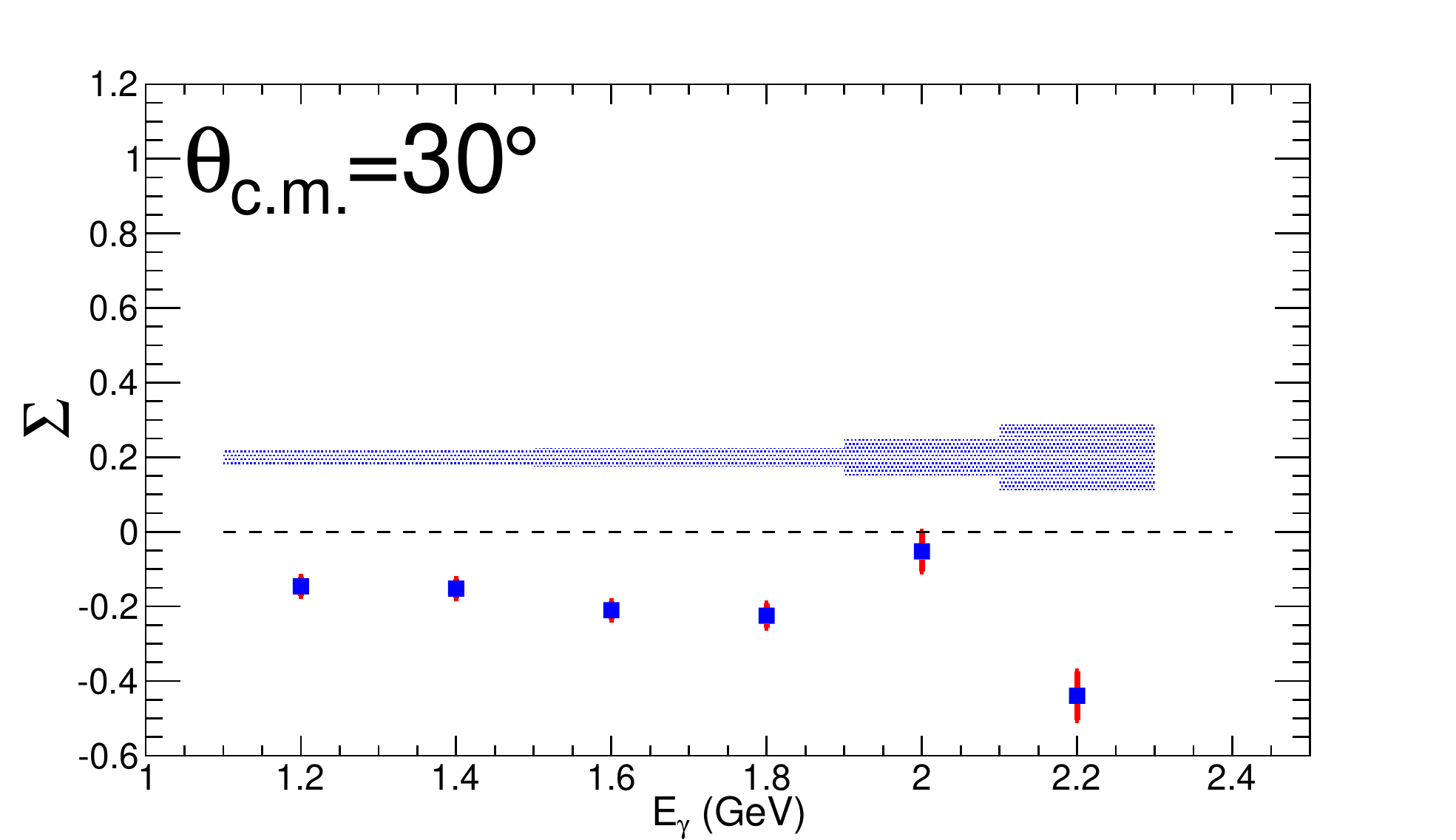} 
   \includegraphics[ width=3.0 in]{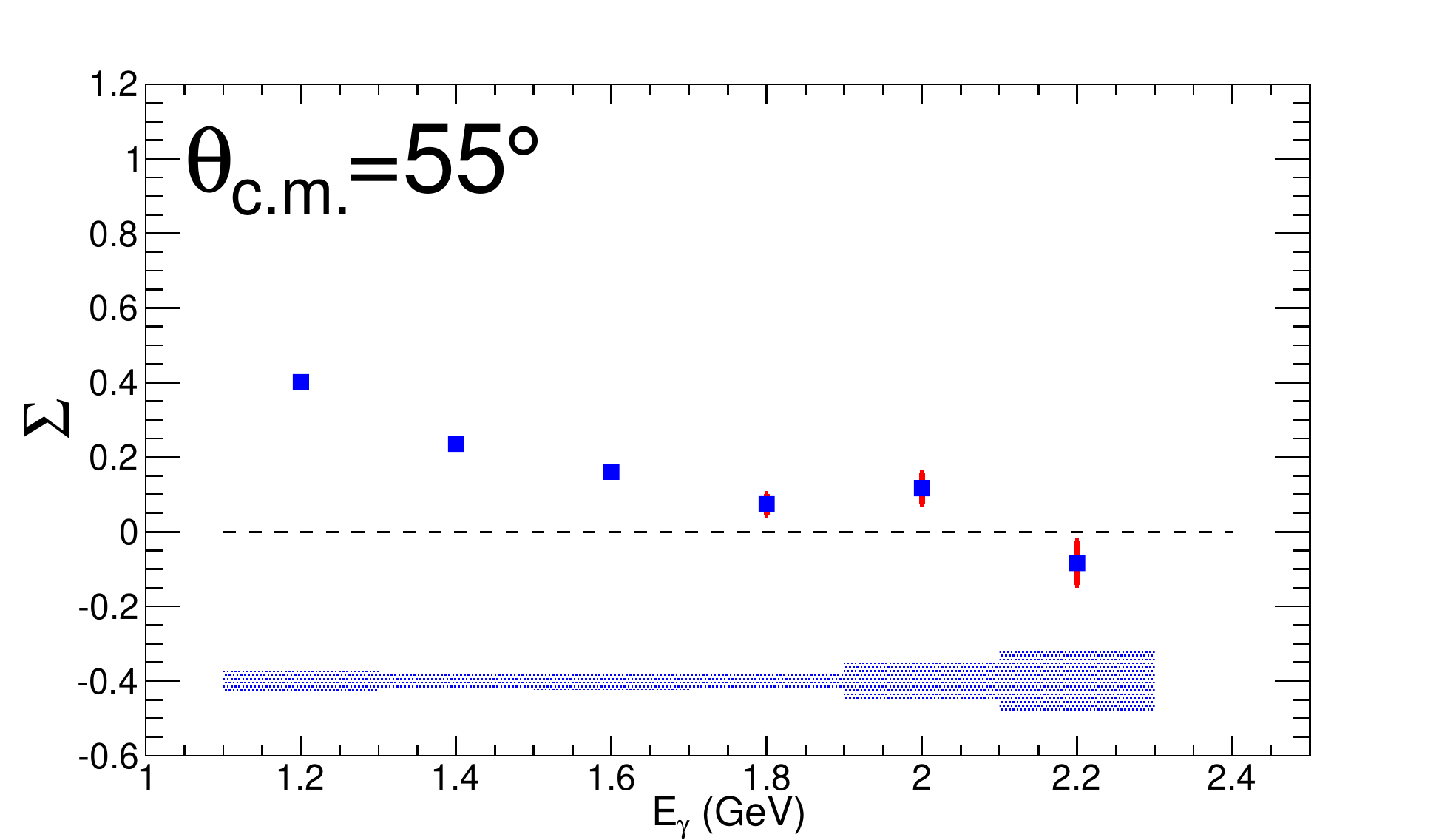} 
   \includegraphics[ width=3.0 in]{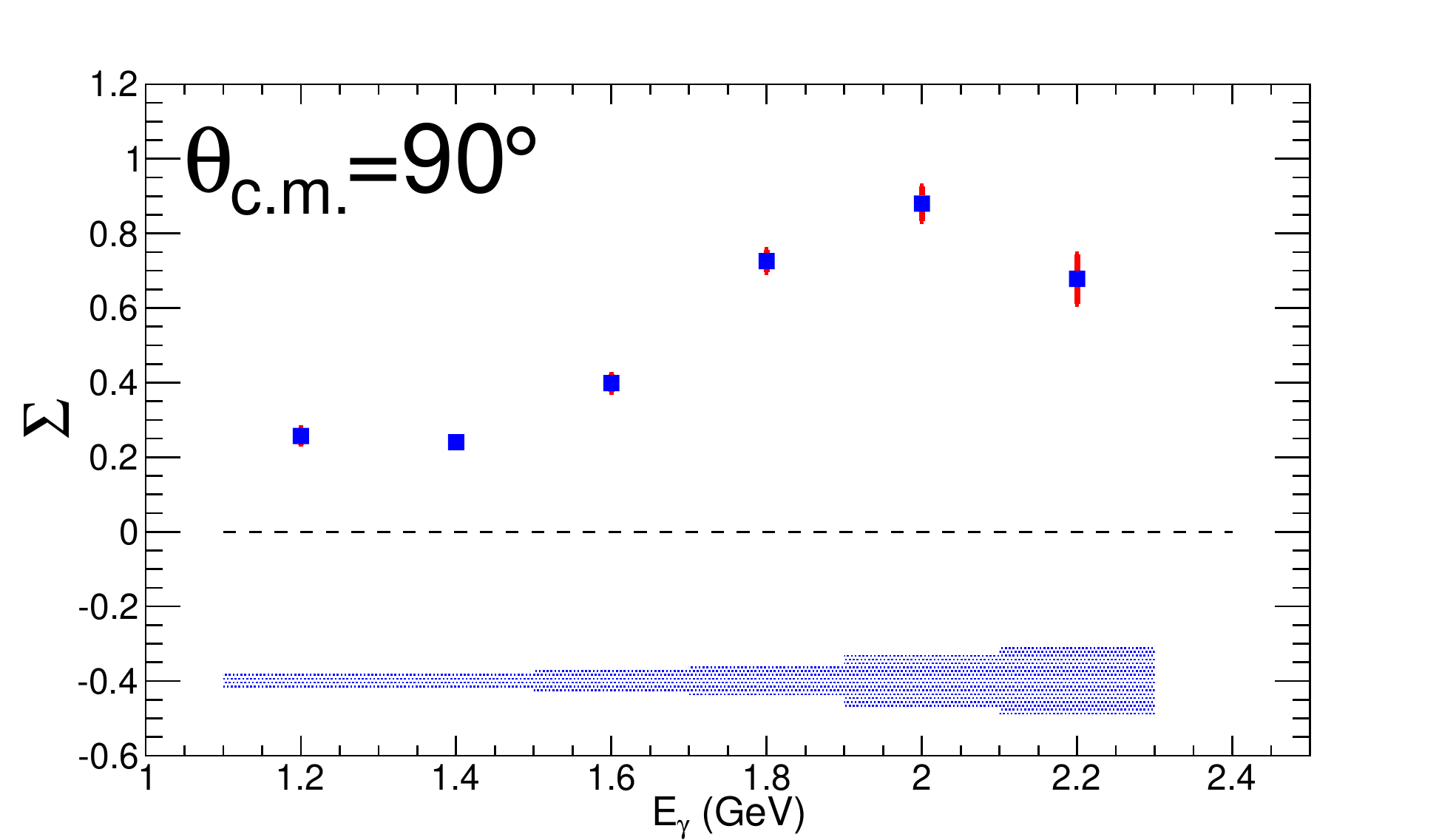} 
   \includegraphics[ width=3.0 in]{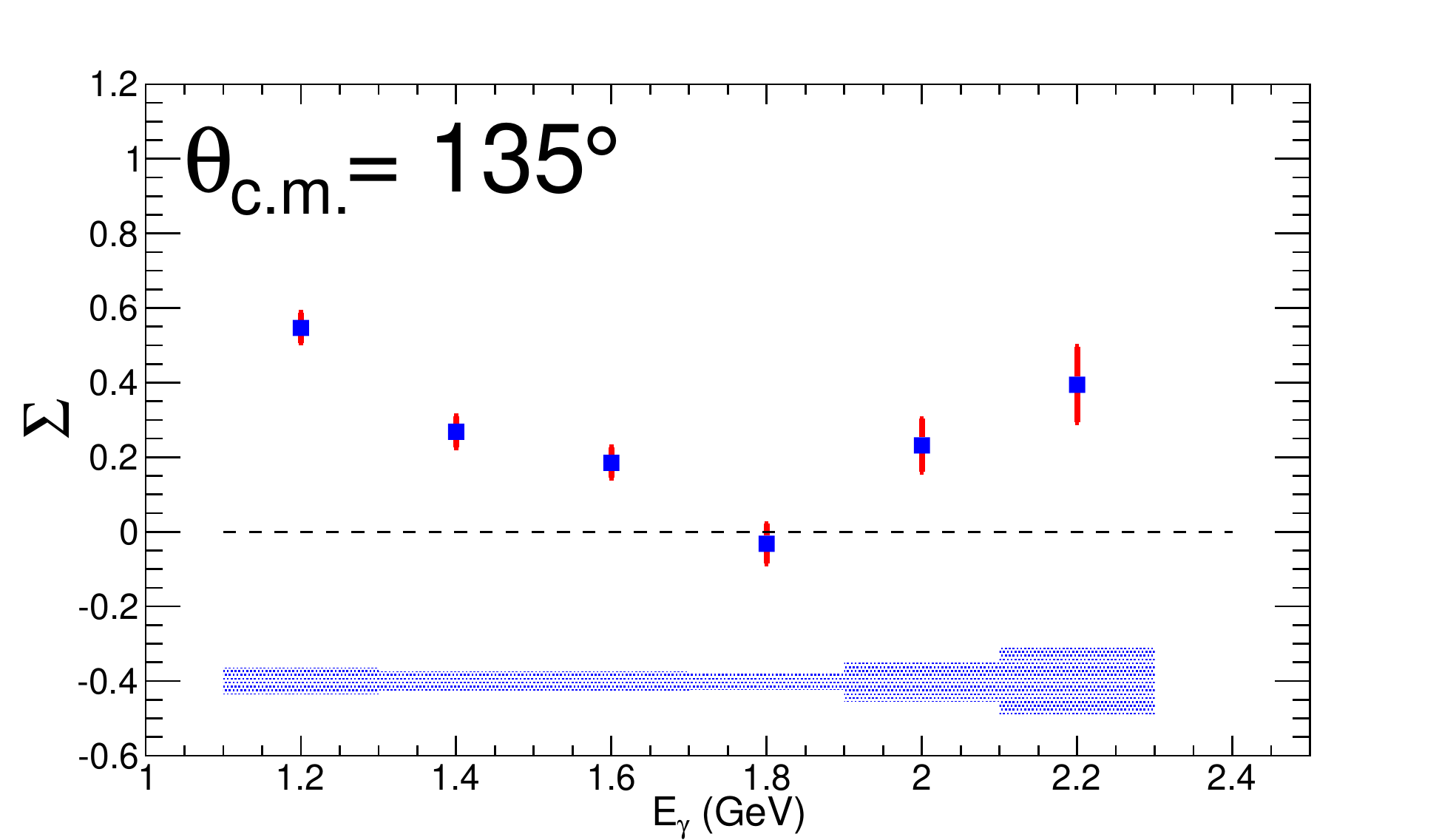} 
   \caption{(Color online) Beam-spin asymmetry as a function of photon energy for four different proton angles in the center-of-mass frame. The angular bin size varies with photon energy. The blue bands indicate the systematic uncertainties of the present CLAS measurements.}      \label{NewResultsFig2a}
   \end{center}
\end{figure*}


\subsection{Angular distributions}\label{Adistributions}

Figure~\ref{NewResultsFig2aaa} shows the angular dependence of $\Sigma$ for the six photon-energy bins  (200-MeV wide) between $E_\gamma=$1.1 and $E_\gamma=$2.3 GeV, as well as the QGSM and HRM predictions. The width of the angular bin varies in an attempt to have constant statistical uncertainties. The results indicate that the beam-spin asymmetry has a local minimum at $\theta_{c.m.}=90^\circ$ for the lowest photon-energy bins. This minimum evolves to a maximum  for the higher photon-energy bins. The data exhibit, especially at the three lower energies, complex structures. At small angles the observable is negative, but increases with the production angle and reaches a positive maximum below $\theta_{c.m.}=90^\circ$. Then, it decreases to a positive minimum shortly above $\theta_{c.m.}=90^\circ$, and reaches a second maximum at large angles. As the photon energy increases, the position of the first maximum shifts towards $\theta_{c.m.}=90^\circ$, while the magnitude of the second maximum, observed at large angles, continuously decreases and $\Sigma$ becomes negative at these large angles for $E_\gamma$ above 1.7 GeV. While the QGSM model predicts similarly complex angular distributions, there are significant differences between the data and the model. The latter predicts positive $\Sigma$ at all energies and angles, while data show that in some bins $\Sigma$ is negative. The positive maximum at small angles in the model is not confirmed by the data. The positive maximum at large angles is confirmed only by the lowest-energy data, but at a different angle.
 The QGSM seems to predict well the maxima at $\theta_{c.m.}=90^\circ$, observed in the higher photon energy data. 
 The discrepancies between the data and the QGSM may be due to resonance contributions to the reaction dynamics, which are not well described in the model.
The HRM model is expected to be valid only for  $\theta_{c.m.}=90^\circ$ angles, where both $t_N$ and $u_N$ are large. Nevertheless, predictions of the HRM model are shown to demonstrate the kinematics where the data exclude applicability of the model.
\begin{figure*}[!]
   \begin{center}
   \includegraphics[ width=6.3 in]{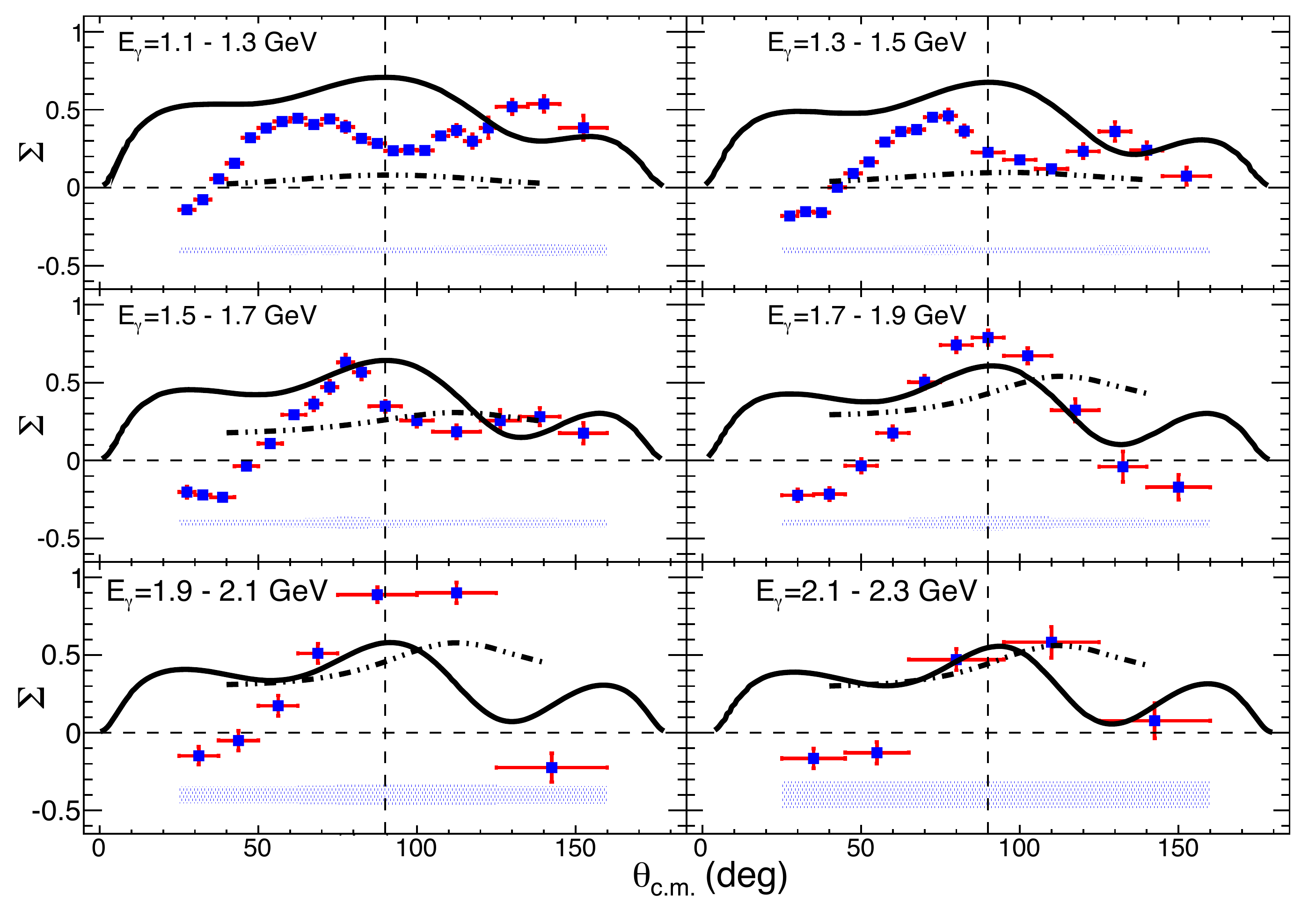} 
   \caption{(Color online) Beam-spin asymmetry as a function of the proton center-of-mass angle, $\theta_{c.m.}$, for the six photon-energy bins between 1.1 and 2.3 GeV. The solid and dash-dotted lines are the QGSM~\cite{Deute38} and HRM~\cite{Deute37, Deute37a} predictions, respectively. The blue bands indicate the systematic uncertainties of the present CLAS measurements.}      \label{NewResultsFig2aaa}
   \end{center}
\end{figure*}

\subsection{Conclusions}\label{GenRem}
The work presented here increases significantly the kinematical coverage and precision of the available data for the beam-spin asymmetry, $\Sigma$, of the reaction $\gamma d\rightarrow pn$.
State-of-the-art models have limited success in reproducing the details of the data. The fact that the models capture only some of the most general features of the data suggests that improvements are needed in the details of the reaction dynamics.  For example, a better phenomenological input to the HRM for the spin dependence of the elementary $pn$ amplitudes could bring the calculation closer to the data. Alternatively, the comparison at $\theta_{c.m.}=90^\circ$ may suggest that the energy range of our data are below the full applicability of the model. 
 The CLAS data provide stringent constraints that can be used in the development of the existing models or even aid in the development of new phenomenological approaches that attempt to describe the underlying dynamics in the transition region from hadronic to partonic degrees of freedom.

\begin{acknowledgments}
First and  foremost we would like to express our gratitude to the late Prof. B. L. Berman for his help and support of this project. His contribution to this work was invaluable and we are grateful for his involvement. 
We would also like to acknowledge the outstanding efforts of the staff of the Accelerator
and the Physics Divisions at Jefferson Lab that made this experiment possible. 

This work was supported in part by the U.S. Department of Energy
and National Science Foundation, the French Centre
National de la Recherche Scientifique and Commissariat
 \`a l'Energie Atomique, the French-American Cultural
 Exchange (FACE), the Italian Istituto Nazionale di
 Fisica Nucleare, the Chilean Comisi\`on Nacional de
 Investigaci\'on Cient\'ifica y Tecnol\'ogica (CONICYT),
 the National Research Foundation of Korea, the State Committee of Science of Armenia, and the
 UK Science and Technology Facilities Council (STFC).
 The Jefferson Science Associates (JSA) operates the
 Thomas Jefferson National Accelerator Facility for the
 United States Department of Energy under contract
 DE-AC05-06OR23177.

\end{acknowledgments}

\cleardoublepage
\bibliography{references.bib}

\begin{thebibliography}{74}%
\makeatletter
\providecommand \@ifxundefined [1]{%
 \@ifx{#1\undefined}
}%
\providecommand \@ifnum [1]{%
 \ifnum #1\expandafter \@firstoftwo
 \else \expandafter \@secondoftwo
 \fi
}%
\providecommand \@ifx [1]{%
 \ifx #1\expandafter \@firstoftwo
 \else \expandafter \@secondoftwo
 \fi
}%
\providecommand \natexlab [1]{#1}%
\providecommand \enquote  [1]{``#1''}%
\providecommand \bibnamefont  [1]{#1}%
\providecommand \bibfnamefont [1]{#1}%
\providecommand \citenamefont [1]{#1}%
\providecommand \href@noop [0]{\@secondoftwo}%
\providecommand \href [0]{\begingroup \@sanitize@url \@href}%
\providecommand \@href[1]{\@@startlink{#1}\@@href}%
\providecommand \@@href[1]{\endgroup#1\@@endlink}%
\providecommand \@sanitize@url [0]{\catcode `\\12\catcode `\$12\catcode
  `\&12\catcode `\#12\catcode `\^12\catcode `\_12\catcode `\%12\relax}%
\providecommand \@@startlink[1]{}%
\providecommand \@@endlink[0]{}%
\providecommand \url  [0]{\begingroup\@sanitize@url \@url }%
\providecommand \@url [1]{\endgroup\@href {#1}{\urlprefix }}%
\providecommand \urlprefix  [0]{URL }%
\providecommand \Eprint [0]{\href }%
\providecommand \doibase [0]{http://dx.doi.org/}%
\providecommand \selectlanguage [0]{\@gobble}%
\providecommand \bibinfo  [0]{\@secondoftwo}%
\providecommand \bibfield  [0]{\@secondoftwo}%
\providecommand \translation [1]{[#1]}%
\providecommand \BibitemOpen [0]{}%
\providecommand \bibitemStop [0]{}%
\providecommand \bibitemNoStop [0]{.\EOS\space}%
\providecommand \EOS [0]{\spacefactor3000\relax}%
\providecommand \BibitemShut  [1]{\csname bibitem#1\endcsname}%
\let\auto@bib@innerbib\@empty
\bibitem [{\citenamefont {Napolitano{\it{ et al.}}}(1988)}]{Deute01}%
  \BibitemOpen
  \bibfield  {author} {\bibinfo {author} {\bibfnamefont {J.}~\bibnamefont
  {Napolitano{\it{ et al.}}}},\ }\href@noop {} {\bibfield  {journal} {\bibinfo
  {journal} {Phys. Rev. Lett.}\ }\textbf {\bibinfo {volume} {61}},\ \bibinfo
  {pages} {2530} (\bibinfo {year} {1988})}\BibitemShut {NoStop}%
\bibitem [{\citenamefont {Freedman{\it{ et al.}}}(1993)}]{Deute02}%
  \BibitemOpen
  \bibfield  {author} {\bibinfo {author} {\bibfnamefont {S.}~\bibnamefont
  {Freedman{\it{ et al.}}}},\ }\href@noop {} {\bibfield  {journal} {\bibinfo
  {journal} {Phys. Rev. C}\ }\textbf {\bibinfo {volume} {48}},\ \bibinfo
  {pages} {1864} (\bibinfo {year} {1993})}\BibitemShut {NoStop}%
\bibitem [{\citenamefont {Belz{\it{ et al.}}}(1995)}]{Deute03}%
  \BibitemOpen
  \bibfield  {author} {\bibinfo {author} {\bibfnamefont {J.~E.}\ \bibnamefont
  {Belz{\it{ et al.}}}},\ }\href@noop {} {\bibfield  {journal} {\bibinfo
  {journal} {Phys. Rev. Lett.}\ }\textbf {\bibinfo {volume} {74}},\ \bibinfo
  {pages} {646} (\bibinfo {year} {1995})}\BibitemShut {NoStop}%
\bibitem [{\citenamefont {Crawford{\it{ et al.}}}(1996)}]{Deute04}%
  \BibitemOpen
  \bibfield  {author} {\bibinfo {author} {\bibfnamefont {R.}~\bibnamefont
  {Crawford{\it{ et al.}}}},\ }\href@noop {} {\bibfield  {journal} {\bibinfo
  {journal} {Nucl. Phys. A}\ }\textbf {\bibinfo {volume} {603}},\ \bibinfo
  {pages} {303} (\bibinfo {year} {1996})}\BibitemShut {NoStop}%
\bibitem [{\citenamefont {Bochna{\it{ et al.}}}(1998)}]{Deute05}%
  \BibitemOpen
  \bibfield  {author} {\bibinfo {author} {\bibfnamefont {C.}~\bibnamefont
  {Bochna{\it{ et al.}}}},\ }\href@noop {} {\bibfield  {journal} {\bibinfo
  {journal} {Phys. Rev. Lett.}\ }\textbf {\bibinfo {volume} {81}},\ \bibinfo
  {pages} {4576} (\bibinfo {year} {1998})}\BibitemShut {NoStop}%
\bibitem [{\citenamefont {Schulte{\it{ et al.}}}(2001)}]{Deute06}%
  \BibitemOpen
  \bibfield  {author} {\bibinfo {author} {\bibfnamefont {E.~C.}\ \bibnamefont
  {Schulte{\it{ et al.}}}},\ }\href@noop {} {\bibfield  {journal} {\bibinfo
  {journal} {Phys. Rev. Lett.}\ }\textbf {\bibinfo {volume} {87}},\ \bibinfo
  {pages} {102302} (\bibinfo {year} {2001})}\BibitemShut {NoStop}%
\bibitem [{\citenamefont {Schulte{\it{ et al.}}}(2002)}]{Deute06a}%
  \BibitemOpen
  \bibfield  {author} {\bibinfo {author} {\bibfnamefont {E.~C.}\ \bibnamefont
  {Schulte{\it{ et al.}}}},\ }\href@noop {} {\bibfield  {journal} {\bibinfo
  {journal} {Phys. Rev. C}\ }\textbf {\bibinfo {volume} {66}},\ \bibinfo
  {pages} {042201R} (\bibinfo {year} {2002})}\BibitemShut {NoStop}%
\bibitem [{\citenamefont {Mirazita{\it{ et al.}}}(2004)}]{Deute09}%
  \BibitemOpen
  \bibfield  {author} {\bibinfo {author} {\bibfnamefont {M.}~\bibnamefont
  {Mirazita{\it{ et al.}}}},\ }\href@noop {} {\bibfield  {journal} {\bibinfo
  {journal} {Phys. Rev. C}\ }\textbf {\bibinfo {volume} {70}},\ \bibinfo
  {pages} {014005} (\bibinfo {year} {2004})}\BibitemShut {NoStop}%
\bibitem [{\citenamefont {Wijesooriya{\it{ et al.}}}(2001)}]{Deute07}%
  \BibitemOpen
  \bibfield  {author} {\bibinfo {author} {\bibfnamefont {K.}~\bibnamefont
  {Wijesooriya{\it{ et al.}}}},\ }\href@noop {} {\bibfield  {journal} {\bibinfo
   {journal} {Phys. Rev. Lett.}\ }\textbf {\bibinfo {volume} {86}},\ \bibinfo
  {pages} {2975} (\bibinfo {year} {2001})}\BibitemShut {NoStop}%
\bibitem [{\citenamefont {Jiang{\it{ et al.}}}(2007)}]{PhysRevLett.98.182302}%
  \BibitemOpen
  \bibfield  {author} {\bibinfo {author} {\bibfnamefont {X.}~\bibnamefont
  {Jiang{\it{ et al.}}}},\ }\href {\doibase 10.1103/PhysRevLett.98.182302}
  {\bibfield  {journal} {\bibinfo  {journal} {Phys. Rev. Lett.}\ }\textbf
  {\bibinfo {volume} {98}},\ \bibinfo {pages} {182302} (\bibinfo {year}
  {2007})}\BibitemShut {NoStop}%
\bibitem [{\citenamefont {Adamian{\it{ et al.}}}(1991)}]{Deute11}%
  \BibitemOpen
  \bibfield  {author} {\bibinfo {author} {\bibfnamefont {F.}~\bibnamefont
  {Adamian{\it{ et al.}}}},\ }\href@noop {} {\bibfield  {journal} {\bibinfo
  {journal} {J. Phys. G}\ }\textbf {\bibinfo {volume} {17}},\ \bibinfo {pages}
  {1189} (\bibinfo {year} {1991})}\BibitemShut {NoStop}%
\bibitem [{\citenamefont {Adamian{\it{ et al.}}}(2000)}]{Deute12}%
  \BibitemOpen
  \bibfield  {author} {\bibinfo {author} {\bibfnamefont {F.}~\bibnamefont
  {Adamian{\it{ et al.}}}},\ }\href@noop {} {\bibfield  {journal} {\bibinfo
  {journal} {Eur. Phys. J. A}\ }\textbf {\bibinfo {volume} {8}},\ \bibinfo
  {pages} {423} (\bibinfo {year} {2000})}\BibitemShut {NoStop}%
\bibitem [{\citenamefont {Rossi{\it{ et al.}}}(2005)}]{Deute13}%
  \BibitemOpen
  \bibfield  {author} {\bibinfo {author} {\bibfnamefont {P.}~\bibnamefont
  {Rossi{\it{ et al.}}}},\ }\href@noop {} {\bibfield  {journal} {\bibinfo
  {journal} {Phys. Rev. Lett.}\ }\textbf {\bibinfo {volume} {94}},\ \bibinfo
  {pages} {012301} (\bibinfo {year} {2005})}\BibitemShut {NoStop}%
\bibitem [{\citenamefont {Matveev}\ \emph {et~al.}(1973)\citenamefont
  {Matveev}, \citenamefont {Muradyan},\ and\ \citenamefont
  {Tavkhelidze}}]{Deute17a}%
  \BibitemOpen
  \bibfield  {author} {\bibinfo {author} {\bibfnamefont {V.}~\bibnamefont
  {Matveev}}, \bibinfo {author} {\bibfnamefont {R.}~\bibnamefont {Muradyan}}, \
  and\ \bibinfo {author} {\bibfnamefont {A.~N.}\ \bibnamefont {Tavkhelidze}},\
  }\href@noop {} {\bibfield  {journal} {\bibinfo  {journal} {Lett. Nuovo
  Cimento}\ }\textbf {\bibinfo {volume} {7}},\ \bibinfo {pages} {719} (\bibinfo
  {year} {1973})}\BibitemShut {NoStop}%
\bibitem [{\citenamefont {Brodsky}\ and\ \citenamefont
  {Farrar}(1973)}]{Deute17}%
  \BibitemOpen
  \bibfield  {author} {\bibinfo {author} {\bibfnamefont {S.~J.}\ \bibnamefont
  {Brodsky}}\ and\ \bibinfo {author} {\bibfnamefont {G.}~\bibnamefont
  {Farrar}},\ }\href@noop {} {\bibfield  {journal} {\bibinfo  {journal} {Phys.
  Rev. Lett.}\ }\textbf {\bibinfo {volume} {31}},\ \bibinfo {pages} {1153}
  (\bibinfo {year} {1973})}\BibitemShut {NoStop}%
\bibitem [{\citenamefont {Lepage}\ and\ \citenamefont
  {Brodsky}(1980)}]{Lepage}%
  \BibitemOpen
  \bibfield  {author} {\bibinfo {author} {\bibfnamefont {G.}~\bibnamefont
  {Lepage}}\ and\ \bibinfo {author} {\bibfnamefont {S.~J.}\ \bibnamefont
  {Brodsky}},\ }\href@noop {} {\bibfield  {journal} {\bibinfo  {journal} {Phys.
  Rev. D}\ }\textbf {\bibinfo {volume} {22}},\ \bibinfo {pages} {2157}
  (\bibinfo {year} {1980})}\BibitemShut {NoStop}%
\bibitem [{\citenamefont {Polchinski}\ and\ \citenamefont
  {Strassler}(2002)}]{cft1}%
  \BibitemOpen
  \bibfield  {author} {\bibinfo {author} {\bibfnamefont {J.}~\bibnamefont
  {Polchinski}}\ and\ \bibinfo {author} {\bibfnamefont {M.}~\bibnamefont
  {Strassler}},\ }\href@noop {} {\bibfield  {journal} {\bibinfo  {journal}
  {Phys. Rev. Lett.}\ }\textbf {\bibinfo {volume} {88}},\ \bibinfo {pages}
  {031601} (\bibinfo {year} {2002})}\BibitemShut {NoStop}%
\bibitem [{\citenamefont {Bower}\ and\ \citenamefont {Tan}(2003)}]{cft2}%
  \BibitemOpen
  \bibfield  {author} {\bibinfo {author} {\bibfnamefont {R.~C.}\ \bibnamefont
  {Bower}}\ and\ \bibinfo {author} {\bibfnamefont {C.~I.}\ \bibnamefont
  {Tan}},\ }\href@noop {} {\bibfield  {journal} {\bibinfo  {journal} {Nucl.
  Phys. B}\ }\textbf {\bibinfo {volume} {662}},\ \bibinfo {pages} {393}
  (\bibinfo {year} {2003})}\BibitemShut {NoStop}%
\bibitem [{\citenamefont {Adreev}(2003)}]{cft3}%
  \BibitemOpen
  \bibfield  {author} {\bibinfo {author} {\bibfnamefont {O.}~\bibnamefont
  {Adreev}},\ }\href@noop {} {\bibfield  {journal} {\bibinfo  {journal} {Phys.
  Rev. D}\ }\textbf {\bibinfo {volume} {67}},\ \bibinfo {pages} {046001}
  (\bibinfo {year} {2003})}\BibitemShut {NoStop}%
\bibitem [{\citenamefont {Maldacena}(1998)}]{maldacena}%
  \BibitemOpen
  \bibfield  {author} {\bibinfo {author} {\bibfnamefont {J.~M.}\ \bibnamefont
  {Maldacena}},\ }\href@noop {} {\bibfield  {journal} {\bibinfo  {journal}
  {Adv. Theor. Math. Phys.}\ }\textbf {\bibinfo {volume} {2}},\ \bibinfo
  {pages} {231} (\bibinfo {year} {1998})}\BibitemShut {NoStop}%
\bibitem [{\citenamefont {Anderson{\it{ et al.}}}(1976)}]{diff_cross}%
  \BibitemOpen
  \bibfield  {author} {\bibinfo {author} {\bibfnamefont {R.~L.}\ \bibnamefont
  {Anderson{\it{ et al.}}}},\ }\href@noop {} {\bibfield  {journal} {\bibinfo
  {journal} {Phys. Rev. D}\ }\textbf {\bibinfo {volume} {14}},\ \bibinfo
  {pages} {679} (\bibinfo {year} {1976})}\BibitemShut {NoStop}%
\bibitem [{\citenamefont {White{\it{ et al.}}}(1994)}]{white}%
  \BibitemOpen
  \bibfield  {author} {\bibinfo {author} {\bibfnamefont {C.}~\bibnamefont
  {White{\it{ et al.}}}},\ }\href@noop {} {\bibfield  {journal} {\bibinfo
  {journal} {Phys. Rev. D}\ }\textbf {\bibinfo {volume} {49}},\ \bibinfo
  {pages} {58} (\bibinfo {year} {1994})}\BibitemShut {NoStop}%
\bibitem [{\citenamefont {Holt}\ and\ \citenamefont
  {Gilman}(2012)}]{holtgilman}%
  \BibitemOpen
  \bibfield  {author} {\bibinfo {author} {\bibfnamefont {R.}~\bibnamefont
  {Holt}}\ and\ \bibinfo {author} {\bibfnamefont {R.}~\bibnamefont {Gilman}},\
  }\href@noop {} {\bibfield  {journal} {\bibinfo  {journal} {Rept. Prog.
  Phys.}\ }\textbf {\bibinfo {volume} {75}},\ \bibinfo {pages} {086301}
  (\bibinfo {year} {2012})}\BibitemShut {NoStop}%
\bibitem [{\citenamefont {Zhu{\it{ et al.}}}(2003)}]{zhu1}%
  \BibitemOpen
  \bibfield  {author} {\bibinfo {author} {\bibfnamefont {L.~Y.}\ \bibnamefont
  {Zhu{\it{ et al.}}}},\ }\href@noop {} {\bibfield  {journal} {\bibinfo
  {journal} {Phys. Rev. Lett.}\ }\textbf {\bibinfo {volume} {91}},\ \bibinfo
  {pages} {022003} (\bibinfo {year} {2003})}\BibitemShut {NoStop}%
\bibitem [{\citenamefont {Zhu{\it{ et al.}}}(2005)}]{zhu2}%
  \BibitemOpen
  \bibfield  {author} {\bibinfo {author} {\bibfnamefont {L.~Y.}\ \bibnamefont
  {Zhu{\it{ et al.}}}},\ }\href@noop {} {\bibfield  {journal} {\bibinfo
  {journal} {Phys. Rev. C}\ }\textbf {\bibinfo {volume} {71}},\ \bibinfo
  {pages} {044603} (\bibinfo {year} {2005})}\BibitemShut {NoStop}%
\bibitem [{\citenamefont {Akerlof{\it{ et al.}}}(1967)}]{p-p1}%
  \BibitemOpen
  \bibfield  {author} {\bibinfo {author} {\bibfnamefont {C.~W.}\ \bibnamefont
  {Akerlof{\it{ et al.}}}},\ }\href@noop {} {\bibfield  {journal} {\bibinfo
  {journal} {Phys. Rev.}\ }\textbf {\bibinfo {volume} {159}},\ \bibinfo {pages}
  {1138} (\bibinfo {year} {1967})}\BibitemShut {NoStop}%
\bibitem [{\citenamefont {Kammerud{\it{ et al.}}}(1971)}]{p-p1a}%
  \BibitemOpen
  \bibfield  {author} {\bibinfo {author} {\bibfnamefont {R.}~\bibnamefont
  {Kammerud{\it{ et al.}}}},\ }\href@noop {} {\bibfield  {journal} {\bibinfo
  {journal} {Phys. Rev. D}\ }\textbf {\bibinfo {volume} {4}},\ \bibinfo {pages}
  {1309} (\bibinfo {year} {1971})}\BibitemShut {NoStop}%
\bibitem [{\citenamefont {Jenkins{\it{ et al.}}}(1978)}]{p-p1b}%
  \BibitemOpen
  \bibfield  {author} {\bibinfo {author} {\bibfnamefont {K.~A.}\ \bibnamefont
  {Jenkins{\it{ et al.}}}},\ }\href@noop {} {\bibfield  {journal} {\bibinfo
  {journal} {Phys. Rev. Lett.}\ }\textbf {\bibinfo {volume} {40}},\ \bibinfo
  {pages} {425} (\bibinfo {year} {1978})}\BibitemShut {NoStop}%
\bibitem [{\citenamefont {Hendry}(1974)}]{p-p2}%
  \BibitemOpen
  \bibfield  {author} {\bibinfo {author} {\bibfnamefont {A.~W.}\ \bibnamefont
  {Hendry}},\ }\href@noop {} {\bibfield  {journal} {\bibinfo  {journal} {Phys.
  Rev. D}\ }\textbf {\bibinfo {volume} {10}},\ \bibinfo {pages} {2300}
  (\bibinfo {year} {1974})}\BibitemShut {NoStop}%
\bibitem [{\citenamefont {Owen{\it{ et al.}}}(1969)}]{pi-p}%
  \BibitemOpen
  \bibfield  {author} {\bibinfo {author} {\bibfnamefont {D.~P.}\ \bibnamefont
  {Owen{\it{ et al.}}}},\ }\href@noop {} {\bibfield  {journal} {\bibinfo
  {journal} {Phys. Rev.}\ }\textbf {\bibinfo {volume} {181}},\ \bibinfo {pages}
  {1794} (\bibinfo {year} {1969})}\BibitemShut {NoStop}%
\bibitem [{\citenamefont {Jenkins{\it{ et al.}}}(1980)}]{pi-pa}%
  \BibitemOpen
  \bibfield  {author} {\bibinfo {author} {\bibfnamefont {K.~A.}\ \bibnamefont
  {Jenkins{\it{ et al.}}}},\ }\href@noop {} {\bibfield  {journal} {\bibinfo
  {journal} {Phys. Rev. D}\ }\textbf {\bibinfo {volume} {21}},\ \bibinfo
  {pages} {2445} (\bibinfo {year} {1980})}\BibitemShut {NoStop}%
\bibitem [{\citenamefont {Baglin{\it{ et al.}}}(1983)}]{pi-pb}%
  \BibitemOpen
  \bibfield  {author} {\bibinfo {author} {\bibfnamefont {C.}~\bibnamefont
  {Baglin{\it{ et al.}}}},\ }\href@noop {} {\bibfield  {journal} {\bibinfo
  {journal} {Nucl. Phys. B}\ }\textbf {\bibinfo {volume} {216}},\ \bibinfo
  {pages} {1} (\bibinfo {year} {1983})}\BibitemShut {NoStop}%
\bibitem [{\citenamefont {Meekins{\it{ et al.}}}(1999)}]{meekins}%
  \BibitemOpen
  \bibfield  {author} {\bibinfo {author} {\bibfnamefont {D.~G.}\ \bibnamefont
  {Meekins{\it{ et al.}}}},\ }\href@noop {} {\bibfield  {journal} {\bibinfo
  {journal} {Phys. Rev. C}\ }\textbf {\bibinfo {volume} {60}},\ \bibinfo
  {pages} {052201} (\bibinfo {year} {1999})}\BibitemShut {NoStop}%
\bibitem [{\citenamefont {Ilieva{\it{ et al.}}}(2006)}]{ilieva}%
  \BibitemOpen
  \bibfield  {author} {\bibinfo {author} {\bibfnamefont {Y.}~\bibnamefont
  {Ilieva{\it{ et al.}}}},\ }\href@noop {} {\bibfield  {journal} {\bibinfo
  {journal} {AIP Conf. Proc.}\ }\textbf {\bibinfo {volume} {842}},\ \bibinfo
  {pages} {431} (\bibinfo {year} {2006})}\BibitemShut {NoStop}%
\bibitem [{\citenamefont {Barannik{\it{ et al.}}}(1986)}]{HelcAmpl}%
  \BibitemOpen
  \bibfield  {author} {\bibinfo {author} {\bibfnamefont {V.}~\bibnamefont
  {Barannik{\it{ et al.}}}},\ }\href@noop {} {\bibfield  {journal} {\bibinfo
  {journal} {Nuclear Physics A}\ }\textbf {\bibinfo {volume} {451}},\ \bibinfo
  {pages} {751 } (\bibinfo {year} {1986})}\BibitemShut {NoStop}%
\bibitem [{\citenamefont {Brodsky}\ and\ \citenamefont
  {Hiller}(1983)}]{Deute21}%
  \BibitemOpen
  \bibfield  {author} {\bibinfo {author} {\bibfnamefont {S.}~\bibnamefont
  {Brodsky}}\ and\ \bibinfo {author} {\bibfnamefont {J.}~\bibnamefont
  {Hiller}},\ }\href@noop {} {\bibfield  {journal} {\bibinfo  {journal} {Phys.
  Rev. C}\ }\textbf {\bibinfo {volume} {28}},\ \bibinfo {pages} {475} (\bibinfo
  {year} {1983})}\BibitemShut {NoStop}%
\bibitem [{\citenamefont {Brodsky}\ \emph {et~al.}(2001)\citenamefont
  {Brodsky}, \citenamefont {Hiller}, \citenamefont {Ji},\ and\ \citenamefont
  {Miller}}]{Deute21a}%
  \BibitemOpen
  \bibfield  {author} {\bibinfo {author} {\bibfnamefont {S.}~\bibnamefont
  {Brodsky}}, \bibinfo {author} {\bibfnamefont {J.}~\bibnamefont {Hiller}},
  \bibinfo {author} {\bibfnamefont {C.-R.}\ \bibnamefont {Ji}}, \ and\ \bibinfo
  {author} {\bibfnamefont {G.~A.}\ \bibnamefont {Miller}},\ }\href@noop {}
  {\bibfield  {journal} {\bibinfo  {journal} {Phys. Rev. C}\ }\textbf {\bibinfo
  {volume} {64}},\ \bibinfo {pages} {055204} (\bibinfo {year}
  {2001})}\BibitemShut {NoStop}%
\bibitem [{\citenamefont {Frankfurt}\ \emph
  {et~al.}(2000{\natexlab{a}})\citenamefont {Frankfurt}, \citenamefont
  {Miller}, \citenamefont {Sargsian},\ and\ \citenamefont
  {Strikman}}]{Deute24}%
  \BibitemOpen
  \bibfield  {author} {\bibinfo {author} {\bibfnamefont {L.}~\bibnamefont
  {Frankfurt}}, \bibinfo {author} {\bibfnamefont {G.}~\bibnamefont {Miller}},
  \bibinfo {author} {\bibfnamefont {M.~M.}\ \bibnamefont {Sargsian}}, \ and\
  \bibinfo {author} {\bibfnamefont {M.}~\bibnamefont {Strikman}},\ }\href@noop
  {} {\bibfield  {journal} {\bibinfo  {journal} {Phys. Rev. Lett.}\ }\textbf
  {\bibinfo {volume} {84}},\ \bibinfo {pages} {3045} (\bibinfo {year}
  {2000}{\natexlab{a}})}\BibitemShut {NoStop}%
\bibitem [{\citenamefont {Frankfurt}\ \emph
  {et~al.}(2000{\natexlab{b}})\citenamefont {Frankfurt}, \citenamefont
  {Miller}, \citenamefont {Sargsian},\ and\ \citenamefont
  {Strikman}}]{Deute25}%
  \BibitemOpen
  \bibfield  {author} {\bibinfo {author} {\bibfnamefont {L.}~\bibnamefont
  {Frankfurt}}, \bibinfo {author} {\bibfnamefont {G.}~\bibnamefont {Miller}},
  \bibinfo {author} {\bibfnamefont {M.}~\bibnamefont {Sargsian}}, \ and\
  \bibinfo {author} {\bibfnamefont {M.}~\bibnamefont {Strikman}},\ }\href@noop
  {} {\bibfield  {journal} {\bibinfo  {journal} {Nucl. Phys. A}\ }\textbf
  {\bibinfo {volume} {663}},\ \bibinfo {pages} {349} (\bibinfo {year}
  {2000}{\natexlab{b}})}\BibitemShut {NoStop}%
\bibitem [{\citenamefont {Julia-Diaz}\ and\ \citenamefont
  {Lee}(2003)}]{Deute25a}%
  \BibitemOpen
  \bibfield  {author} {\bibinfo {author} {\bibfnamefont {B.}~\bibnamefont
  {Julia-Diaz}}\ and\ \bibinfo {author} {\bibfnamefont {T.-S.~H.}\ \bibnamefont
  {Lee}},\ }\href@noop {} {\bibfield  {journal} {\bibinfo  {journal} {Mod.
  Phys. Lett. A}\ }\textbf {\bibinfo {volume} {18}},\ \bibinfo {pages} {200}
  (\bibinfo {year} {2003})}\BibitemShut {NoStop}%
\bibitem [{\citenamefont {Sargsian}\ and\ \citenamefont
  {Granados}(2009)}]{misakgranados}%
  \BibitemOpen
  \bibfield  {author} {\bibinfo {author} {\bibfnamefont {M.~M.}\ \bibnamefont
  {Sargsian}}\ and\ \bibinfo {author} {\bibfnamefont {C.~G.}\ \bibnamefont
  {Granados}},\ }\href@noop {} {\bibfield  {journal} {\bibinfo  {journal}
  {Phys. Rev. C}\ }\textbf {\bibinfo {volume} {80}},\ \bibinfo {pages} {014612}
  (\bibinfo {year} {2009})}\BibitemShut {NoStop}%
\bibitem [{\citenamefont {Kondratyuk{\it{ et al.}}}(1993)}]{Deute22}%
  \BibitemOpen
  \bibfield  {author} {\bibinfo {author} {\bibfnamefont {L.}~\bibnamefont
  {Kondratyuk{\it{ et al.}}}},\ }\href@noop {} {\bibfield  {journal} {\bibinfo
  {journal} {Phys. Rev. C}\ }\textbf {\bibinfo {volume} {48}},\ \bibinfo
  {pages} {2491} (\bibinfo {year} {1993})}\BibitemShut {NoStop}%
\bibitem [{\citenamefont {Grishina{\it{ et al.}}}(2001)}]{Deute23}%
  \BibitemOpen
  \bibfield  {author} {\bibinfo {author} {\bibfnamefont {V.~Y.}\ \bibnamefont
  {Grishina{\it{ et al.}}}},\ }\href@noop {} {\bibfield  {journal} {\bibinfo
  {journal} {Eur. Phys. J. A}\ }\textbf {\bibinfo {volume} {10}},\ \bibinfo
  {pages} {355} (\bibinfo {year} {2001})}\BibitemShut {NoStop}%
\bibitem [{\citenamefont {Brodsky}\ \emph {et~al.}(2004)\citenamefont
  {Brodsky}, \citenamefont {Frankfurt}, \citenamefont {Gilman}, \citenamefont
  {Hiller}, \citenamefont {Miller}, \citenamefont {Piasetzky}, \citenamefont
  {Sargsian},\ and\ \citenamefont {Strikman}}]{RNA-rev}%
  \BibitemOpen
  \bibfield  {author} {\bibinfo {author} {\bibfnamefont {S.~J.}\ \bibnamefont
  {Brodsky}}, \bibinfo {author} {\bibfnamefont {L.~L.}\ \bibnamefont
  {Frankfurt}}, \bibinfo {author} {\bibfnamefont {R.}~\bibnamefont {Gilman}},
  \bibinfo {author} {\bibfnamefont {J.~R.}\ \bibnamefont {Hiller}}, \bibinfo
  {author} {\bibfnamefont {G.~A.}\ \bibnamefont {Miller}}, \bibinfo {author}
  {\bibfnamefont {E.}~\bibnamefont {Piasetzky}}, \bibinfo {author}
  {\bibfnamefont {M.}~\bibnamefont {Sargsian}}, \ and\ \bibinfo {author}
  {\bibfnamefont {M.~I.}\ \bibnamefont {Strikman}},\ }\href@noop {} {\bibfield
  {journal} {\bibinfo  {journal} {Phys. Lett. B}\ }\textbf {\bibinfo {volume}
  {578}},\ \bibinfo {pages} {69} (\bibinfo {year} {2004})}\BibitemShut
  {NoStop}%
\bibitem [{\citenamefont {Pomerantz{\it{ et al.}}}(2010)}]{pomerantz}%
  \BibitemOpen
  \bibfield  {author} {\bibinfo {author} {\bibfnamefont {I.}~\bibnamefont
  {Pomerantz{\it{ et al.}}}},\ }\href@noop {} {\bibfield  {journal} {\bibinfo
  {journal} {Phys. Rev. B}\ }\textbf {\bibinfo {volume} {684}},\ \bibinfo
  {pages} {106} (\bibinfo {year} {2010})}\BibitemShut {NoStop}%
\bibitem [{\citenamefont {Granados}\ and\ \citenamefont
  {Sargsian}(2009)}]{misakgranados2}%
  \BibitemOpen
  \bibfield  {author} {\bibinfo {author} {\bibfnamefont {C.~G.}\ \bibnamefont
  {Granados}}\ and\ \bibinfo {author} {\bibfnamefont {M.~M.}\ \bibnamefont
  {Sargsian}},\ }\href@noop {} {\bibfield  {journal} {\bibinfo  {journal}
  {Phys. Rev. Lett.}\ }\textbf {\bibinfo {volume} {103}},\ \bibinfo {pages}
  {212001} (\bibinfo {year} {2009})}\BibitemShut {NoStop}%
\bibitem [{\citenamefont {{M.~L.~Perl, J.~Cox, M.~J.~Longo and
  M.~Kreisler}}(1970)}]{dataforAmplitude1}%
  \BibitemOpen
  \bibfield  {author} {\bibinfo {author} {\bibnamefont {{M.~L.~Perl, J.~Cox,
  M.~J.~Longo and M.~Kreisler}}},\ }\href@noop {} {\bibfield  {journal}
  {\bibinfo  {journal} {Phys. Rev. D}\ }\textbf {\bibinfo {volume} {1}},\
  \bibinfo {pages} {1857} (\bibinfo {year} {1970})}\BibitemShut {NoStop}%
\bibitem [{\citenamefont {{J.~L.~Stone, J.~P.~Chanowski, H.~R.~Gustafson,
  M.~J.~Longo and S.~W.~Gray}}(1978)}]{dataforAmplitude2}%
  \BibitemOpen
  \bibfield  {author} {\bibinfo {author} {\bibnamefont {{J.~L.~Stone,
  J.~P.~Chanowski, H.~R.~Gustafson, M.~J.~Longo and S.~W.~Gray}}},\ }\href@noop
  {} {\bibfield  {journal} {\bibinfo  {journal} {Nucl. Phys. B}\ }\textbf
  {\bibinfo {volume} {143}},\ \bibinfo {pages} {1} (\bibinfo {year}
  {1978})}\BibitemShut {NoStop}%
\bibitem [{\citenamefont {Kaidalov}(1982)}]{kaidalov}%
  \BibitemOpen
  \bibfield  {author} {\bibinfo {author} {\bibfnamefont {A.~B.}\ \bibnamefont
  {Kaidalov}},\ }\href@noop {} {\bibfield  {journal} {\bibinfo  {journal} {Z.
  Phys. C}\ }\textbf {\bibinfo {volume} {12}},\ \bibinfo {pages} {63} (\bibinfo
  {year} {1982})}\BibitemShut {NoStop}%
\bibitem [{\citenamefont {Kaidalov}(1999)}]{kaidalova}%
  \BibitemOpen
  \bibfield  {author} {\bibinfo {author} {\bibfnamefont {A.~B.}\ \bibnamefont
  {Kaidalov}},\ }\href@noop {} {\bibfield  {journal} {\bibinfo  {journal}
  {Surv. High Energy Phys.}\ }\textbf {\bibinfo {volume} {13}},\ \bibinfo
  {pages} {265} (\bibinfo {year} {1999})}\BibitemShut {NoStop}%
\bibitem [{\citenamefont {Arakelian}\ \emph {et~al.}(1994)\citenamefont
  {Arakelian}, \citenamefont {Grigorian}, \citenamefont {Ivanov},\ and\
  \citenamefont {Kaidalov}}]{QGSMspin1}%
  \BibitemOpen
  \bibfield  {author} {\bibinfo {author} {\bibfnamefont {G.~G.}\ \bibnamefont
  {Arakelian}}, \bibinfo {author} {\bibfnamefont {A.~A.}\ \bibnamefont
  {Grigorian}}, \bibinfo {author} {\bibfnamefont {N.~Y.}\ \bibnamefont
  {Ivanov}}, \ and\ \bibinfo {author} {\bibfnamefont {A.~B.}\ \bibnamefont
  {Kaidalov}},\ }\href@noop {} {\bibfield  {journal} {\bibinfo  {journal} {Z.
  Phys. C}\ }\textbf {\bibinfo {volume} {{\bf 63}}},\ \bibinfo {pages} {137}
  (\bibinfo {year} {1994})}\BibitemShut {NoStop}%
\bibitem [{\citenamefont {Ananikian}\ and\ \citenamefont
  {Ivanov}(2007)}]{QGSMspin2}%
  \BibitemOpen
  \bibfield  {author} {\bibinfo {author} {\bibfnamefont {L.~N.}\ \bibnamefont
  {Ananikian}}\ and\ \bibinfo {author} {\bibfnamefont {N.~Y.}\ \bibnamefont
  {Ivanov}},\ }\href@noop {} {\bibfield  {journal} {\bibinfo  {journal} {Phys.\
  Rev.\ D}\ }\textbf {\bibinfo {volume} {{\bf 75}}},\ \bibinfo {pages} {014010}
  (\bibinfo {year} {2007})}\BibitemShut {NoStop}%
\bibitem [{\citenamefont {Grishina{\it{ et al.}}}(2004)}]{Deute38}%
  \BibitemOpen
  \bibfield  {author} {\bibinfo {author} {\bibfnamefont {V.~Y.}\ \bibnamefont
  {Grishina{\it{ et al.}}}},\ }\href@noop {} {\bibfield  {journal} {\bibinfo
  {journal} {Eur. Phys. J. A}\ }\textbf {\bibinfo {volume} {19}},\ \bibinfo
  {pages} {117} (\bibinfo {year} {2004})}\BibitemShut {NoStop}%
\bibitem [{\citenamefont {Sargsian}(2004)}]{Deute37}%
  \BibitemOpen
  \bibfield  {author} {\bibinfo {author} {\bibfnamefont {M.}~\bibnamefont
  {Sargsian}},\ }\href@noop {} {\bibfield  {journal} {\bibinfo  {journal}
  {Phys. Lett. B}\ }\textbf {\bibinfo {volume} {587}},\ \bibinfo {pages} {41}
  (\bibinfo {year} {2004})}\BibitemShut {NoStop}%
\bibitem [{\citenamefont {Sargsian}()}]{Deute37a}%
  \BibitemOpen
  \bibfield  {author} {\bibinfo {author} {\bibfnamefont {M.}~\bibnamefont
  {Sargsian}},\ }\href@noop {} {\bibinfo  {journal} {Private Communication}\
  }\BibitemShut {NoStop}%
\bibitem [{\citenamefont {Mecking{\it{ et al.}}}(2003)}]{RefExp2}%
  \BibitemOpen
\bibfield  {journal} {  }\bibfield  {author} {\bibinfo {author} {\bibfnamefont
  {B.~A.}\ \bibnamefont {Mecking{\it{ et al.}}}},\ }\href@noop {} {\bibfield
  {journal} {\bibinfo  {journal} {Nucl. Instr. and Meth. A}\ }\textbf {\bibinfo
  {volume} {503}},\ \bibinfo {pages} {513} (\bibinfo {year}
  {2003})}\BibitemShut {NoStop}%
\bibitem [{\citenamefont {Mestayer{\it{ et al.}}}(2000)}]{RefExp11}%
  \BibitemOpen
  \bibfield  {author} {\bibinfo {author} {\bibfnamefont {M.~D.}\ \bibnamefont
  {Mestayer{\it{ et al.}}}},\ }\href@noop {} {\bibfield  {journal} {\bibinfo
  {journal} {Nucl. Instr. and Meth. A}\ }\textbf {\bibinfo {volume} {449}},\
  \bibinfo {pages} {81} (\bibinfo {year} {2000})}\BibitemShut {NoStop}%
\bibitem [{\citenamefont {Smith{\it{ et al.}}}(1999)}]{RefExp15}%
  \BibitemOpen
  \bibfield  {author} {\bibinfo {author} {\bibfnamefont {E.~S.}\ \bibnamefont
  {Smith{\it{ et al.}}}},\ }\href@noop {} {\bibfield  {journal} {\bibinfo
  {journal} {Nucl. Instr. and Meth. A}\ }\textbf {\bibinfo {volume} {\bf432}},\
  \bibinfo {pages} {265} (\bibinfo {year} {1999})}\BibitemShut {NoStop}%
\bibitem [{\citenamefont {Adams{\it{ et al.}}}(2001)}]{RefExp14}%
  \BibitemOpen
  \bibfield  {author} {\bibinfo {author} {\bibfnamefont {G.}~\bibnamefont
  {Adams{\it{ et al.}}}},\ }\href@noop {} {\bibfield  {journal} {\bibinfo
  {journal} {Nucl. Instr. and Meth. A}\ }\textbf {\bibinfo {volume} {\bf465}},\
  \bibinfo {pages} {414} (\bibinfo {year} {2001})}\BibitemShut {NoStop}%
\bibitem [{\citenamefont {Amarian{\it{ et al.}}}(2001)}]{RefExp16}%
  \BibitemOpen
  \bibfield  {author} {\bibinfo {author} {\bibfnamefont {M.}~\bibnamefont
  {Amarian{\it{ et al.}}}},\ }\href@noop {} {\bibfield  {journal} {\bibinfo
  {journal} {Nucl. Instr. and Meth. A}\ }\textbf {\bibinfo {volume} {\bf460}},\
  \bibinfo {pages} {239} (\bibinfo {year} {2001})}\BibitemShut {NoStop}%
\bibitem [{\citenamefont {Taylor{\it{ et al.}}}(2001)}]{RefExp16a}%
  \BibitemOpen
  \bibfield  {author} {\bibinfo {author} {\bibfnamefont {S.}~\bibnamefont
  {Taylor{\it{ et al.}}}},\ }\href@noop {} {\bibfield  {journal} {\bibinfo
  {journal} {Nucl. Instr. and Meth. A}\ }\textbf {\bibinfo {volume} {\bf462}},\
  \bibinfo {pages} {484} (\bibinfo {year} {2001})}\BibitemShut {NoStop}%
\bibitem [{\citenamefont {Livingston}()}]{KenLivingCLAS2011}%
  \BibitemOpen
  \bibfield  {author} {\bibinfo {author} {\bibfnamefont {K.}~\bibnamefont
  {Livingston}},\ }\href@noop {} {\ \textbf {\bibinfo {volume} {CLAS Note
  2011-020}},\ \bibinfo {pages} {Jefferson Laboratory}}\BibitemShut {NoStop}%
\bibitem [{\citenamefont {Sober{\it{ et al.}}}(2000)}]{RefExp10}%
  \BibitemOpen
  \bibfield  {author} {\bibinfo {author} {\bibfnamefont {D.~I.}\ \bibnamefont
  {Sober{\it{ et al.}}}},\ }\href@noop {} {\bibfield  {journal} {\bibinfo
  {journal} {Nucl. Instr. and Meth. A}\ }\textbf {\bibinfo {volume} {440}},\
  \bibinfo {pages} {263} (\bibinfo {year} {2000})}\BibitemShut {NoStop}%
\bibitem [{\citenamefont {Nadel-Turonski{\it{ et al.}}}(2006)}]{g13exp}%
  \BibitemOpen
  \bibfield  {author} {\bibinfo {author} {\bibfnamefont {P.}~\bibnamefont
  {Nadel-Turonski{\it{ et al.}}}},\ }\href@noop {} {\bibfield  {journal}
  {\bibinfo  {journal} {JLab Approved Experiments}\ }\textbf {\bibinfo {volume}
  {E-06-103}} (\bibinfo {year} {2006})}\BibitemShut {NoStop}%
\bibitem [{\citenamefont {Pasyuk}()}]{RefAnaI5}%
  \BibitemOpen
  \bibfield  {author} {\bibinfo {author} {\bibfnamefont {E.}~\bibnamefont
  {Pasyuk}},\ }\href@noop {} {\ \textbf {\bibinfo {volume} {CLAS Note
  2007-016}},\ \bibinfo {pages} {Jefferson Laboratory}}\BibitemShut {NoStop}%
\bibitem [{\citenamefont {Williams}\ \emph {et~al.}()\citenamefont {Williams},
  \citenamefont {Applegate},\ and\ \citenamefont {Meyer}}]{RefAnaI6}%
  \BibitemOpen
  \bibfield  {author} {\bibinfo {author} {\bibfnamefont {M.}~\bibnamefont
  {Williams}}, \bibinfo {author} {\bibfnamefont {D.}~\bibnamefont {Applegate}},
  \ and\ \bibinfo {author} {\bibfnamefont {C.}~\bibnamefont {Meyer}},\
  }\href@noop {} {\ \textbf {\bibinfo {volume} {CLAS Note 2004-017}},\ \bibinfo
  {pages} {Jefferson Laboratory}}\BibitemShut {NoStop}%
\bibitem [{\citenamefont {Mattione}(2007)}]{RefAnaI7}%
  \BibitemOpen
  \bibfield  {author} {\bibinfo {author} {\bibfnamefont {P.}~\bibnamefont
  {Mattione}},\ }\href@noop {} {\ \textbf {\bibinfo {volume} {{\it Kinematic
  Fitting of Detached Vertices}, M.S. Thesis, Rice University}} (\bibinfo
  {year} {2007})},\ \bibinfo {note}
  {\url{http://www1.jlab.org/Ul/Publications/view\_pub.cfm?pub\_id=7379}}\BibitemShut
  {NoStop}%
\bibitem [{\citenamefont {Williams}(2007)}]{MwillProb1}%
  \BibitemOpen
  \bibfield  {author} {\bibinfo {author} {\bibfnamefont {M.}~\bibnamefont
  {Williams}},\ }\href@noop {} {\ \textbf {\bibinfo {volume} {{\it Measurement
  of Differential Cross Sections and Spin Density Matrix Elements along with a
  Partial Wave Analysis for $\gamma p\rightarrow p \omega$ using CLAS at
  Jefferson Lab}, Ph.D. Thesis, Carnegie Mellon University}} (\bibinfo {year}
  {2007})},\ \bibinfo {note}
  {\url{www.jlab.org/Hall-B/general/clas_thesis.html}}\BibitemShut {NoStop}%
\bibitem [{\citenamefont {Williams}\ \emph {et~al.}(2009)\citenamefont
  {Williams}, \citenamefont {Bellis},\ and\ \citenamefont {Meyer}}]{RefAnaI8}%
  \BibitemOpen
  \bibfield  {author} {\bibinfo {author} {\bibfnamefont {M.}~\bibnamefont
  {Williams}}, \bibinfo {author} {\bibfnamefont {M.}~\bibnamefont {Bellis}}, \
  and\ \bibinfo {author} {\bibfnamefont {C.}~\bibnamefont {Meyer}},\
  }\href@noop {} {\ \textbf {\bibinfo {volume} {JINST 4}},\ \bibinfo {pages}
  {P10003} (\bibinfo {year} {2009})}\BibitemShut {NoStop}%
\bibitem [{\citenamefont {\"{U}berall}(1962)}]{RefAnaII3b}%
  \BibitemOpen
  \bibfield  {author} {\bibinfo {author} {\bibfnamefont {H.}~\bibnamefont
  {\"{U}berall}},\ }\href@noop {} {\bibfield  {journal} {\bibinfo  {journal}
  {Z. Naturforsch}\ }\textbf {\bibinfo {volume} {17a}},\ \bibinfo {pages} {332}
  (\bibinfo {year} {1962})}\BibitemShut {NoStop}%
\bibitem [{\citenamefont {Zachariou}(2012)}]{nickthesis}%
  \BibitemOpen
  \bibfield  {author} {\bibinfo {author} {\bibfnamefont {N.}~\bibnamefont
  {Zachariou}},\ }\href@noop {} {\ \textbf {\bibinfo {volume} {{\it
  Determination of the Azimuthal Asymmetry of Deuteron Photodisintegration in
  the Energy Region $E_\gamma=1.1-2.3~$GeV}, Ph.D. Thesis, George Washington
  University}} (\bibinfo {year} {2012})},\ \bibinfo {note}
  {\url{http://www.jlab.org/Hall-B/general/thesis/Zachariou_thesis.pdf}}\BibitemShut
  {NoStop}%
\bibitem [{\citenamefont {Zachariou}\ and\ \citenamefont
  {Ilieva}()}]{NickCLASNOTE}%
  \BibitemOpen
  \bibfield  {author} {\bibinfo {author} {\bibfnamefont {N.}~\bibnamefont
  {Zachariou}}\ and\ \bibinfo {author} {\bibfnamefont {Y.}~\bibnamefont
  {Ilieva}},\ }\href@noop {} {\ \textbf {\bibinfo {volume} {CLAS Note
  2012-011}},\ \bibinfo {pages} {Jefferson Laboratory}}\BibitemShut {NoStop}%
\bibitem [{\citenamefont {Sokhan}(2009)}]{RefIntro3}%
  \BibitemOpen
  \bibfield  {author} {\bibinfo {author} {\bibfnamefont {D.}~\bibnamefont
  {Sokhan}},\ }\href@noop {} {\ \textbf {\bibinfo {volume} {{\it Beam Asymmetry
  Measurement from Pion Photoproduction on the Neutron}, Ph.D. Thesis,
  University of Edinburg}} (\bibinfo {year} {2009})},\ \bibinfo {note}
  {\url{https://www.jlab.org/Hall-B/general/thesis/Sokhan_thesis.pdf}}\BibitemShut
  {NoStop}%
\bibitem [{\citenamefont {Dugger}\ and\ \citenamefont
  {Ritchie}()}]{DuggerCLASNOTE}%
  \BibitemOpen
  \bibfield  {author} {\bibinfo {author} {\bibfnamefont {M.}~\bibnamefont
  {Dugger}}\ and\ \bibinfo {author} {\bibfnamefont {B.~G.}\ \bibnamefont
  {Ritchie}},\ }\href@noop {} {\ \textbf {\bibinfo {volume} {CLAS Note
  2012-002}},\ \bibinfo {pages} {Jefferson Laboratory}}\BibitemShut {NoStop}%
\end{thebibliography}%

\end{document}